# Percolation and Connectivity in the Intrinsically Secure Communications Graph

Pedro C. Pinto, *Student Member, IEEE*, and Moe Z. Win, *Fellow, IEEE*

Corresponding Address:

Pedro C. Pinto

Laboratory for Information and Decision Systems (LIDS)

Massachusetts Institute of Technology (MIT)

77 Massachusetts Avenue, Room 32-D674

Cambridge, MA 02139 USA

Tel.: (857) 928-6444

e-mail: `ppinto@alum.mit.edu`

P. C. Pinto and M. Z. Win are with the Laboratory for Information and Decision Systems (LIDS), Massachusetts Institute of Technology, Room 32-D674, 77 Massachusetts Avenue, Cambridge, MA 02139, USA (e-mail: `ppinto@alum.mit.edu`, `moewin@mit.edu`).

This research was supported, in part, by the Portuguese Science and Technology Foundation under grant SFRH-BD-17388-2004; the MIT Institute for Soldier Nanotechnologies; the Office of Naval Research under Presidential Early Career Award for Scientists and Engineers (PECASE) N00014-09-1-0435; and the National Science Foundation under grant ECS-0636519.

CONTENTS

**Abstract**

The ability to exchange secret information is critical to many commercial, governmental, and military networks. The *intrinsically secure communications graph* ($i\mathcal{S}$-graph) is a random graph which describes the connections that can be securely established over a large-scale network, by exploiting the physical properties of the wireless medium. This paper aims to characterize the global properties of the $i\mathcal{S}$-graph in terms of: (i) percolation on the infinite plane, and (ii) full connectivity on a finite region. First, for the Poisson $i\mathcal{S}$-graph defined on the infinite plane, the existence of a phase transition is proven, whereby an unbounded component of connected nodes suddenly arises as the density of legitimate nodes is increased. This shows that long-range secure communication is still possible in the presence of eavesdroppers. Second, full connectivity on a finite region of the Poisson $i\mathcal{S}$-graph is considered. The exact asymptotic behavior of full connectivity in the limit of a large density of legitimate nodes is characterized. Then, simple, explicit expressions are derived in order to closely approximate the probability of full connectivity for a finite density of legitimate nodes. The results help clarify how the presence of eavesdroppers can compromise long-range secure communication.

## I. INTRODUCTION

Contemporary security systems for wireless networks are based on cryptographic primitives that generally ignore two key factors: (a) the physical properties of the underlying communication channels, and (b) the spatial configuration of both the legitimate and malicious nodes. These two factors are important since they affect the propagation channels between the nodes, which in turn determine the fundamental secrecy potential of a wireless network. In fact, the randomness introduced both by the physics of the wireless medium and by the spatial location of the nodes can be leveraged to strengthen the overall security of the communications infrastructure.[1]

The basis for information-theoretic security, which builds on the notion of perfect secrecy [1], was laid in [2] and later in [3], [4]. More recently, there has been a renewed interest in information-theoretic security over wireless channels, from the perspective of space-time communications [5], multiple-input multiple-output communications [6]–[10], eavesdropper collusion [11], [12], cooperative relay networks [13], fading channels [14]–[18], strong secrecy [19], [20], secret key agreement [21]–[25], code design [26]–[28], among other topics. A fundamental limitation of this literature is that it only considers scenarios with a small number of nodes. To account for large-scale networks composed of multiple legitimate and eavesdropper nodes, *secrecy graphs* were introduced in [29] from a geometrical perspective, and in [30] from an information-theoretic perspective. The local connectivity of secrecy graphs was extensively characterized in [31], while the scaling laws of the secrecy capacity were presented in [32], [33].

---

[1] In the literature, the term "security" typically encompasses 3 different characteristics: *secrecy* (or privacy), *integrity*, and *authenticity*. This paper does not consider the issues of integrity or authenticity, and the terms "secrecy" and "security" are used interchangeably.

Percolation theory studies the existence of phase transitions in random graphs, whereby an infinite cluster of connected nodes suddenly arises as some system parameter is varied. Various percolation models have been considered in the literature. The Poisson Boolean model was introduced in [34], which derived the first bounds on the critical density, and started the field of continuum percolation. The Poisson random connection model was introduced and analyzed in [35]. The Poisson nearest neighbour model was considered in [36]. The signal-to-interference-plus-noise ratio (SINR) model was characterized in [37]. A comprehensive study of the various models and results in continuum percolation can be found in [38].

The connectivity of large but finite networks has also received attention the literature. The asymptotic behavior of partial connectivity of the Poisson Boolean model restricted to a finite box was studied in [39]. The asymptotic full connectivity of the same model was analyzed in [40], [41], which obtained the rate of growth of the radius that ensures full connectivity. The asymptotic full connectivity in finite nearest neighbour networks was considered in [42], [43].

In this paper, we characterize long-range secure connectivity in wireless networks by considering the *intrinsically secure communications graph* ($i\mathcal{S}$-graph) as defined in [31]. The $i\mathcal{S}$-graph describes the connections that can be established with information-theoretic security over a large-scale network. We focus on percolation of the $i\mathcal{S}$-graph on the infinite plane, and full connectivity in a finite region. The main contributions of this paper are as follows:

- *Percolation in the $i\mathcal{S}$-graph:* We prove the existence of a phase transition in the Poisson $i\mathcal{S}$-graph defined on the infinite plane, whereby an unbounded component of connected nodes suddenly arises as we increase the density of legitimate nodes. In particular, we determine for which combinations of system parameters does percolation occur. This shows that long-range communication is still possible in a wireless network when a secrecy constraint is present.

- *Full connectivity in the $i\mathcal{S}$-graph:* We analyze secure full connectivity on a finite region of the Poisson $i\mathcal{S}$-graph. We characterize the exact asymptotic behavior of full connectivity in the limit of a large density of legitimate nodes. Then, we obtain simple, explicit expressions that closely approximate the probability of full connectivity for a finite density of legitimate nodes.

This paper is organized as follows. Section II describes the system model. Section III characterizes continuum percolation in the Poisson $i\mathcal{S}$-graph defined over the infinite plane. Section IV analyzes full connectivity in the Poisson $i\mathcal{S}$-graph restricted to a finite region. Section V concludes the paper and summarizes important findings.

## II. SYSTEM MODEL

We start by describing our system model and defining our measures of secrecy. The notation and symbols used throughout the paper are summarized in Table I.

### A. Wireless Propagation Characteristics

In a wireless environment, the received power $P_{\mathrm{rx}}(x_i, x_j)$ associated with the link $\overrightarrow{x_i x_j}$ can modeled as

$$P_{\mathrm{rx}}(x_i, x_j) = P_\ell \cdot g(x_i, x_j, Z_{x_i,x_j}), \tag{1}$$

where $P_\ell$ is the (common) transmit power of the legitimate nodes; and $g(x_i, x_j, Z_{x_i,x_j})$ is the power gain of the link $\overrightarrow{x_i x_j}$, where the random variable (RV) $Z_{x_i,x_j}$ represents the random propagation effects (such as multipath fading or shadowing) associated with link $\overrightarrow{x_i x_j}$. The channel gain $g(x_i, x_j, Z_{x_i,x_j})$ is considered constant (quasi-static) throughout the use of the communications channel, corresponding to channels with a large coherence time. The gain function is assumed to satisfy the following conditions, which are typically observed in practice:

1) $g(x_i, x_j, Z_{x_i,x_j})$ depends on $x_i$ and $x_j$ only through the link length $|x_i - x_j|$; with abuse of notation, we can write $g(r, z) \triangleq g(x_i, x_j, z)|_{|x_i-x_j|\to r}$.
2) $g(r, z)$ is continuous and strictly decreasing in $r$.
3) $\lim_{r\to\infty} g(r, z) = 0$.

The proposed model is general enough to account for common choices of $g$. One example is the unbounded model where $g(r,z) = \frac{z}{r^{2b}}$. The term $\frac{1}{r^{2b}}$ accounts for the far-field path loss with distance, where the amplitude loss exponent $b$ is environment-dependent and can approximately range from $0.8$ (e.g., hallways inside buildings) to $4$ (e.g., dense urban environments), with $b=1$ corresponding to free space propagation. This model is analytically convenient [44], but since the gain becomes unbounded as the distance approaches zero, it must be used with care for extremely dense networks. Another example is the bounded model where $g(r,z) = \frac{z}{1+r^{2b}}$. This model has the same far-field dependence as the unbounded model, but eliminates the singularity at the origin. Unfortunately, it often leads to intractable analytical results. The effect of the singularity at $r=0$ on the performance evaluation of a wireless system is considered in [45].

## B. Wireless Information-Theoretic Security

We now define our measure of secrecy more precisely. While our main interest is targeted towards the behavior of large-scale networks, we briefly review the setup for a single legitimate link with a single eavesdropper. The results therein will serve as basis for the notion of $i\mathcal{S}$-graph to be established later.

Consider the model depicted in Fig. 1, where a legitimate user (Alice) wants to send messages to another user (Bob). Alice encodes a message $s$, represented by a discrete RV, into a codeword, represented by the complex random sequence of length $n$, $x^n = (x(1), \ldots, x(n)) \in \mathbb{C}^n$, for transmission over the channel. Bob observes the output of a discrete-time channel (the *legitimate channel*), which at time $i$ is given by

$$y_\ell(i) = h_\ell \cdot x(i) + n_\ell(i), \quad 1 \leq i \leq n,$$

where $h_\ell \in \mathbb{C}$ is the quasi-static amplitude gain of the legitimate channel,[2] and $n_\ell(i) \sim \mathcal{N}_c(0, \sigma_\ell^2)$ is AWGN with power $\sigma_\ell^2$ per complex sample.[3] Bob makes a decision $\hat{s}$ on $s$ based on the output $y_\ell$, incurring in an error probability equal to $\mathbb{P}\{\hat{s}_\ell \neq s\}$. A third party (Eve) is also capable of eavesdropping on Alice's transmissions. Eve observes the output of a discrete-time channel (the *eavesdropper's channel*), which at time $i$ is given by

$$y_{\text{e}}(i) = h_{\text{e}} \cdot x(i) + n_{\text{e}}(i), \quad 1 \leq i \leq n,$$

where $h_{\text{e}} \in \mathbb{C}$ is the quasi-static amplitude gain of the eavesdropper channel, and $n_{\text{e}}(i) \sim \mathcal{N}_{\text{c}}(0, \sigma_{\text{e}}^2)$ is AWGN with power $\sigma_{\text{e}}^2$ per complex sample. It is assumed that the signals $x$, $h_\ell$, $h_{\text{e}}$, $n_\ell$, and $n_{\text{e}}$ are mutually independent. Each codeword transmitted by Alice is subject to the average power constraint of $P_\ell$ per complex symbol, i.e.,

$$\frac{1}{n} \sum_{i=1}^{n} \mathbb{E}\{|x(i)|^2\} \leq P_\ell. \tag{2}$$

---

[2]The amplitude gain $h_\ell$ can be related to the power gain in (1) as $g(r_\ell, Z_\ell) = |h_\ell|^2$, where $r_\ell$ and $Z_\ell$ are, respectively, the length and random propagation effects of the legitimate link.

[3]We use $\mathcal{N}_c(0, \sigma^2)$ to denote a circularly symmetric (CS) complex Gaussian distribution, where the real and imaginary parts are IID $\mathcal{N}(0, \sigma^2/2)$.

We define the rate of transmission as

$$\mathcal{R} \triangleq \frac{H(s)}{n},$$

where $H(\cdot)$ denotes the entropy function.

Throughout the paper, we use *strong secrecy* as the condition for information-theoretic security, and define it as follows [19].

*Definition 2.1 (Strong Secrecy):* The rate $\mathcal{R}^*$ is said to be *achievable with strong secrecy* if $\forall \epsilon > 0$, for sufficiently large $n$, there exists an encoder-decoder pair with rate $\mathcal{R}$ satisfying the following conditions:

$$\mathcal{R} \geq \mathcal{R}^* - \epsilon,$$

$$H(s|y_{\text{e}}^n) \geq H(s) - \epsilon,$$

$$\mathbb{P}\{\hat{s}_\ell \neq s\} \leq \epsilon.$$

We define the *maximum secrecy rate* (MSR) $\mathcal{R}_{\text{s}}$ of the legitimate channel to be the maximum rate $\mathcal{R}^*$ that is achievable with strong secrecy.[4] If the legitimate link operates at a rate below the MSR $\mathcal{R}_{\text{s}}$, there exists an encoder-decoder pair such that the eavesdropper is unable to obtain additional information about $s$ from the observation $y_{\text{e}}^n$, in the sense that $H(s|y_{\text{e}}^n)$ approaches $H(s)$ as the codeword length $n$ grows. It was shown in [4], [17] that for a given realization of the channel gains $h_\ell, h_e$, the MSR of the Gaussian wiretap channel is

$$\mathcal{R}_s(x_i, x_j) = \left[ \log_2 \left( 1 + \frac{P_\ell \cdot |h_\ell|^2}{\sigma_\ell^2} \right) - \log_2 \left( 1 + \frac{P_\ell \cdot |h_e|^2}{\sigma_e^2} \right) \right]^+, \quad (3)$$

in bits per complex dimension, where $[x] = \max\{x, 0\}$.[5] In the next sections, we use these basic results to analyze secrecy in large-scale networks.

---

[4]See [20] for a comparison between the concepts of weak and strong secrecy. In the case of Gaussian noise, the MSR is *the same* under the weak and strong secrecy conditions.

## C. $i\mathcal{S}$-Graph

Consider a wireless network where the legitimate nodes and the potential eavesdroppers are randomly scattered in space, according to some point processes. The $i\mathcal{S}$-graph is a convenient representation of the information-theoretically secure links that can be established on such network. In the following, we introduce a precise definition of the $i\mathcal{S}$-graph, based on the notion of strong secrecy.

*Definition 2.2 ($i\mathcal{S}$-Graph [31]):* Let $\Pi_\ell = \{x_i\}_{i=1}^\infty \subset \mathbb{R}^d$ denote the set of legitimate nodes, and $\Pi_e = \{e_i\}_{i=1}^\infty \subset \mathbb{R}^d$ denote the set of eavesdroppers. The $i\mathcal{S}$-*graph* is the directed graph $G = \{\Pi_\ell, \mathcal{E}\}$ with vertex set $\Pi_\ell$ and edge set

$$\mathcal{E} = \{\overrightarrow{x_i x_j} : \mathcal{R}_s(x_i, x_j) > \varrho\}, \quad (4)$$

where $\varrho$ is a threshold representing the prescribed infimum secrecy rate for each communication link; and $\mathcal{R}_s(x_i, x_j)$ is the MSR, for a given realization of the channel gains, of the link between the transmitter $x_i$ and the receiver $x_j$, given by

$$\mathcal{R}_s(x_i, x_j) = \left[ \log_2 \left( 1 + \frac{P_{\mathrm{rx}}(x_i, x_j)}{\sigma_\ell^2} \right) - \log_2 \left( 1 + \frac{P_{\mathrm{rx}}(x_i, e^*)}{\sigma_e^2} \right) \right]^+, \quad (5)$$

with

$$e^* = \underset{e_k \in \Pi_e}{\operatorname{argmax}} P_{\mathrm{rx}}(x_i, e_k). \quad (6)$$

This definition presupposes that the eavesdroppers are not allowed to *collude* (i.e., they cannot exchange or combine information), and therefore only the eavesdropper with the strongest received signal from $x_i$ determines the MSR between $x_i$ and $x_j$. The effect of eavesdropper collusion on the local connectivity of the $i\mathcal{S}$-graph is analyzed in [31].

---

[5]Operationally, the MSR $\mathcal{R}_s$ can be achieved if Alice first estimates $h_\ell$ and $h_e$ (i.e., has full CSI), and then uses a code that achieves MSR in the AWGN channel. Estimation of $h_e$ is possible, for instance, when Eve is another active user in the wireless network, so that Alice can estimate the eavesdropper's channel during Eve's transmissions. As we shall see, the $i\mathcal{S}$-graph model presented in this paper relies on an outage formulation, and therefore does *not* require assumptions concerning availability of full CSI.

The $i\mathcal{S}$-graph admits an outage interpretation, in the sense that legitimate nodes set a target secrecy rate $\varrho$ at which they transmit without knowing the channel state information (CSI) of the legitimate nodes and eavesdroppers. In this context, an edge between two nodes signifies that the corresponding channel is not in secrecy outage.

In the remainder of the paper, we consider the case where the following conditions hold: (a) the wireless environment introduces only path loss, i.e., $Z_{x_i,x_j} = 1$ in (1); and (b) the noise powers of the legitimate users and eavesdroppers are equal, i.e., $\sigma_\ell^2 = \sigma_e^2 = \sigma^2$. In such case, we can combine (1), (4), and (5) to obtain the following edge set[6]

$$\mathcal{E} = \left\{ \overrightarrow{x_i x_j} : g(|x_i - x_j|) > 2^\varrho g(|x_i - e^*|) + \frac{\sigma^2}{P}(2^\varrho - 1), \quad e^* = \operatorname*{argmin}_{e_k \in \Pi_e} |x_i - e_k| \right\}, \quad (7)$$

where $e^*$ denotes the eavesdropper closest to the transmitter $x_i$. The particular case of $\varrho = 0$ corresponds to considering the *existence* of secure links, in the sense that an edge $\overrightarrow{x_i x_j}$ is present iff $\mathcal{R}_s(x_i, x_j) > 0$. Thus, a positive (but possibly small) rate exists at which $x_i$ can transmit to $x_j$ with information-theoretic security. In this case, the edge set in (7) simplifies to

$$\mathcal{E} = \left\{ \overrightarrow{x_i x_j} : |x_i - x_j| < |x_i - e^*|, \quad e^* = \operatorname*{argmin}_{e_k \in \Pi_e} |x_i - e_k| \right\}, \quad (8)$$

which corresponds to the geometrical model proposed in [29]. Fig. 2(a) shows an example of such an $i\mathcal{S}$-graph on $\mathbb{R}^2$.

The spatial location of the legitimate and eavesdropper nodes can be modeled either deterministically or stochastically. In many important scenarios, only a statistical description of the node positions is available, and thus a stochastic spatial model is more suitable. In particular, when the node positions are unknown to the network designer a priori, we may as well treat them as completely random according to a homogeneous Poisson point process [46].[7] The Poisson process has maximum entropy among all homogeneous processes [47], and serves as a simple and useful model for the position of nodes in a network [44], [48].

---

[6]For notational simplicity, when $Z = 1$, we omit the second argument of the function $g(r, z)$ and simply use $g(r)$.

[7]The spatial Poisson process is a natural choice in such situation because, given that a node is inside a region $\mathcal{R}$, the PDF of its position is conditionally uniform over $\mathcal{R}$.

*Definition 2.3 (Poisson iS-graph):* The *Poisson iS-graph* is an $i\mathcal{S}$-graph where $\Pi_\ell, \Pi_e \subset \mathbb{R}^d$ are mutually independent, homogeneous Poisson point processes with densities $\lambda_\ell$ and $\lambda_e$, respectively.

In the remainder of the paper (unless otherwise indicated), we focus on Poisson $i\mathcal{S}$-graphs on $\mathbb{R}^2$.

## III. Percolation in the Poisson $i\mathcal{S}$-Graph

Percolation theory studies the behaviour of the connected components in random graphs. Typically, a continuum percolation model consists of an underlying point process defined on the infinite plane, and a rule that describes how connections are established between the nodes [38]. A main property of all percolation models is that they exhibit a *phase transition* as some continuous parameter is varied. If this parameter is the density $\lambda$ of nodes, then the phase transition occurs at some *critical density* $\lambda_c$. When $\lambda < \lambda_c$, denoted as the *subcritical phase*, all the clusters are a.s. bounded.[8] When $\lambda > \lambda_c$, denoted as the *supercritical phase*, the graph exhibits a.s. an unbounded cluster of nodes, or in other words, the graph *percolates*.

Percolation theory plays an important role in the study of connectivity in multi-hop wireless networks, where the formation of an infinite component of connected nodes is desirable for communication over arbitrarily long distances. In the literature, percolation—and therefore long-range communication—was shown to occur in the following models, all of them driven by a Poisson point process:

1) Boolean model [34], where two nodes are directly connected iff their distance is smaller than a fixed radius $r$. This can be used to model unsecured communication subject to a minimum received signal-to-noise ratio (SNR), in the absence of fading.

2) Random connection model [35], where each pair of nodes is directly connected with some probability $p(r)$ depending only on their distance $r$, and independently of every other pair. This can be used to model unsecured communication in the presence of fading, subject to a minimum received SNR.

3) SINR model [37], where two nodes are directly connected if the SINR exceeds the same threshold at both ends. This can be used to model unsecured communication subject to a minimum received SINR, in the absence of fading.

---

[8]We say that an event occurs "almost surely" (a.s.) if its probability is equal to one.

4) Nearest neighbour model [36], where each node connects to its $k$ nearest neighbours. This can be used to model unsecured communication in a centralized network where a power control scheme ensures connectivity to the $k$ nearest nodes only.

In this section, we focus on the $i\mathcal{S}$-graph model, and show that long-range communication with information-theoretic security is feasible in the presence of eavesdroppers. The mathematical characterization of the $i\mathcal{S}$-graph presents two challenges: i) unlike the models 1–4 above, the $i\mathcal{S}$-graph is a directed graph, which leads to the study of *directed percolation*; and ii) similarly to models 3 and 4, the $i\mathcal{S}$-graph exhibits dependencies between the state of different edges, which leads to the study of *dependent percolation*.

In what follows, we start by introducing some definitions, then present and prove the main theorem concerning percolation in the $i\mathcal{S}$-graph, and lastly illustrate the percolation phenomenon with various simulation results. The study of full connectivity in the $i\mathcal{S}$-graph over a finite domain (as opposed to percolation in the infinite plane) is also of interest, and is considered in Section IV.

## A. Definitions

*Graphs:* As before, we use $G = \{\Pi_\ell, \mathcal{E}\}$ to denote the (directed) $i\mathcal{S}$-graph with vertex set $\Pi_\ell$ and edge set given in (4). In addition, we define two undirected graphs: the *weak $i\mathcal{S}$-graph* $G^{\text{weak}} = \{\Pi_\ell, \mathcal{E}^{\text{weak}}\}$, where

$$\mathcal{E}^{\text{weak}} = \{\overline{x_i x_j} : \mathcal{R}_{\mathsf{s}}(x_i, x_j) > \varrho \vee \mathcal{R}_{\mathsf{s}}(x_j, x_i) > \varrho\},$$

and the *strong $i\mathcal{S}$-graph* $G^{\text{strong}} = \{\Pi_\ell, \mathcal{E}^{\text{strong}}\}$, where

$$\mathcal{E}^{\text{strong}} = \{\overline{x_i x_j} : \mathcal{R}_{\mathsf{s}}(x_i, x_j) > \varrho \wedge \mathcal{R}_{\mathsf{s}}(x_j, x_i) > \varrho\}.$$

The graph $G^{\text{weak}}$ represents the links where secure *unidirectional* communication is possible with an MSR greater than $\varrho$. The graph $G^{\text{strong}}$, on the other hand, represents the links where secure *bidirectional* communication is possible with an MSR greater than $\varrho$. The various types of $i\mathcal{S}$-graphs are illustrated in Fig. 2.

*Graph Components:* While the notion of "component" is unambiguous in undirected graphs, some subtleties arise in directed graphs. Specifically, the notion of component is not clear in a directed graph, because even if node $x$ can reach $y$ through a sequence of directed edges, that does not imply that $y$ can reach $x$. We can, however, generalize the notion of component associated with undirected graphs by defining 4 different graph components for the $i\mathcal{S}$-graph.

In what follows, we use the notation $x \xrightarrow{G} y$ to represent a path from node $x$ to node $y$ in a directed graph $G$, and $x \xleftrightarrow{G^*} y$ to represent a path between node $x$ and node $y$ in an undirected graph $G^*$. Let the *out-component* $\mathcal{K}^{\text{out}}(x)$ of node $x$ be the set of nodes which can be reached from node $x$ in the $i\mathcal{S}$-graph $G$, i.e.,

$$\mathcal{K}^{\text{out}}(x) \triangleq \{y \in \Pi_\ell : \exists\, x \xrightarrow{G} y\}. \tag{9}$$

Similarly, let the *in-component* $\mathcal{K}^{\text{in}}$ of node $x$ be the set of nodes from which node $x$ can be reached in the $i\mathcal{S}$-graph $G$, i.e.,

$$\mathcal{K}^{\text{in}}(x) \triangleq \{y \in \Pi_\ell : \exists\, y \xrightarrow{G} x\}. \tag{10}$$

Let the *weak component* $\mathcal{K}^{\text{weak}}(x)$ be the set of nodes which are connected to node $x$ in the weak $i\mathcal{S}$-graph $G^{\text{weak}}$, i.e.,

$$\mathcal{K}^{\text{weak}}(x) \triangleq \{y \in \Pi_\ell : \exists\, x \xleftrightarrow{G^{\text{weak}}} y\}. \tag{11}$$

Let the *strong component* $\mathcal{K}^{\text{strong}}(x)$ be the set of nodes which are connected to node $x$ in the strong $i\mathcal{S}$-graph $G^{\text{strong}}$, i.e.,

$$\mathcal{K}^{\text{strong}}(x) \triangleq \{y \in \Pi_\ell : \exists\, x \xleftrightarrow{G^{\text{strong}}} y\}. \tag{12}$$

From these definitions, it is clear that for a given realization of $\Pi_\ell$ and $\Pi_e$ the following properties hold for any $x$:

$$\mathcal{K}^{\text{strong}}(x) \subseteq \mathcal{K}^{\text{out}}(x) \subseteq \mathcal{K}^{\text{weak}}(x), \tag{13}$$

$$\mathcal{K}^{\text{strong}}(x) \subseteq \mathcal{K}^{\text{in}}(x) \subseteq \mathcal{K}^{\text{weak}}(x), \tag{14}$$

These properties are illustrated in Fig. 9(c).[9]

---

[9]In the literature, the weak and strong components of node $x$ are sometimes defined differently as

$$\mathcal{K}^{\text{weak}}(x) \triangleq \{y \in \Pi : \exists\, x \xrightarrow{G} y \vee \exists\, y \xrightarrow{G} x\} = \mathcal{K}^{\text{out}}(x) \cup \mathcal{K}^{\text{in}}(x),$$

and

$$\mathcal{K}^{\text{strong}}(x) \triangleq \{y \in \Pi : \exists\, x \xrightarrow{G} y \wedge \exists\, y \xrightarrow{G} x\} = \mathcal{K}^{\text{out}}(x) \cap \mathcal{K}^{\text{in}}(x).$$

In this paper, we prefer the definitions in (11) and (12), since they only depend on the respective undirected graphs $G^{\text{weak}}$ and $G^{\text{strong}}$, and do not require explicit knowledge of $G$. As we shall see, this choice will simplify many of the derivations, namely by allowing us to translate an analysis of *directed* graphs into one of *undirected* graphs.

*Percolation Probabilities:* To study the percolation in the $i\mathcal{S}$-graph, it is useful to define percolation probabilities associated with the four graph components. Such probabilities depend on the MSR threshold $\varrho$, and the node densities $\lambda_\ell$ and $\lambda_e$. Specifically, let $p_\infty^{\text{out}}$, $p_\infty^{\text{in}}$, $p_\infty^{\text{weak}}$, and $p_\infty^{\text{strong}}$ respectively be the probabilities that the in, out, weak, and strong components containing node $x = 0$ have an infinite number of nodes, i.e.,[10]

$$p_\infty^\diamond(\lambda_\ell, \lambda_e, \varrho) \triangleq \mathbb{P}\{|\mathcal{K}^\diamond(0)| = \infty\},$$

where $\diamond \in \{\text{out}, \text{in}, \text{weak}, \text{strong}\}$.[11] Our goal is to study the properties and behavior of these percolation probabilities.

## B. Main Result

We now investigate the percolation phenomenon in the $i\mathcal{S}$-graph. Specifically, we aim to determine if percolation in the $i\mathcal{S}$-graph is possible, and if so, for which combinations of system parameters $(\lambda_\ell, \lambda_e, \varrho)$ does it occur. The result is given by the following main theorem.

*Theorem 3.1 (Phase Transition in the $i\mathcal{S}$-Graph):* For any $\lambda_e > 0$ and $\varrho$ satisfying

$$0 \leq \varrho < \varrho_{\max} \triangleq \log_2\left(1 + \frac{P \cdot g(0)}{\sigma^2}\right), \tag{15}$$

there exist critical densities $\lambda_c^{\text{out}}$, $\lambda_c^{\text{in}}$, $\lambda_c^{\text{weak}}$, $\lambda_c^{\text{strong}}$ satisfying

$$0 < \lambda_c^{\text{weak}} \leq \lambda_c^{\text{out}} \leq \lambda_c^{\text{strong}} < \infty \tag{16}$$

$$0 < \lambda_c^{\text{weak}} \leq \lambda_c^{\text{in}} \leq \lambda_c^{\text{strong}} < \infty \tag{17}$$

such that

$$p_\infty^\diamond = 0, \quad \text{for } \lambda_\ell < \lambda_c^\diamond, \tag{18}$$

$$p_\infty^\diamond > 0, \quad \text{for } \lambda_\ell > \lambda_c^\diamond, \tag{19}$$

for any $\diamond \in \{\text{out}, \text{in}, \text{weak}, \text{strong}\}$. Conversely, if $\varrho > \varrho_{\max}$, then $p_\infty^\diamond = 0$ for any $\lambda_\ell, \lambda_e$.

To prove the theorem, we introduce the following three lemmas.

---

[10] We condition on the event that a legitimate node is located at $x = 0$. According to Slivnyak's theorem [49, Sec. 4.4], apart from the given point at $x = 0$, the probabilistic structure of the conditioned process is identical to that of the original process.

[11] Except where otherwise indicated, in the rest of the paper we use the symbol $\diamond$ to represent the out, in, weak, or strong component.

*Lemma 3.1:* For any $\lambda_e > 0$ and $\varrho$ satisfying (15), there exists an $\epsilon > 0$ such that $p_\infty^{\text{weak}}(\lambda_\ell) = 0$ for all $\lambda_\ell < \epsilon$.

*Proof:* Due to its length, the proof is postponed to Section III-C. □

*Lemma 3.2:* For any $\lambda_e > 0$ and $\varrho$ satisfying (15), there exists a $\zeta < \infty$ such that $p_\infty^{\text{strong}}(\lambda_\ell) > 0$ for all $\lambda_\ell > \zeta$.

*Proof:* Due to its length, the proof is postponed to Section III-D. □

*Lemma 3.3:* For any $\lambda_e > 0$ and $\varrho \geq 0$, the percolation probability $p_\infty^\diamond(\lambda_\ell)$ is a non-decreasing function of $\lambda_\ell$.

*Proof:* See Appendix A. □

With these lemmas we are now in condition to prove Theorem 3.1.

*Proof of Theorem 3.1:* We first show that if $\varrho > \varrho_{\max}$, then $p_\infty^\diamond = 0$. The MSR $\mathcal{R}_s$ of a link $\overrightarrow{x_i x_j}$, given in (5), is upper bounded by the channel capacity $\mathcal{R}$ of a link with zero length, i.e., $\mathcal{R}_s(x_i, x_j) \leq \mathcal{R}(x_i, x_i) = \log_2\left(1 + \frac{P \cdot g(0)}{\sigma^2}\right)$. If the threshold $\varrho$ is set such that $\varrho > \varrho_{\max}$, the condition $\mathcal{R}_s(x_i, x_j) > \varrho$ in (4) for the existence of the edge $\overrightarrow{x_i x_j}$ is never satisfied by any $x_i, x_j$. Thus, the $i\mathcal{S}$-graph $G$ has no edges and cannot percolate. We now consider the case of $0 \leq \varrho < \varrho_{\max}$. From properties (13) and (14), we have $p_\infty^{\text{strong}} \leq p_\infty^{\text{out}} \leq p_\infty^{\text{weak}}$ and $p_\infty^{\text{strong}} \leq p_\infty^{\text{in}} \leq p_\infty^{\text{weak}}$. Then, Lemmas 3.1, 3.2, and 3.3 imply that each curve $p_\infty^\diamond(\lambda_\ell)$ departs from the zero value at some critical density $\lambda_c^\diamond$, as expressed by (18) and (19). Furthermore, these critical densities are nontrivial in the sense that $0 < \lambda_c^\diamond < \infty$. The ordering of critical densities in (16) and (17) follows from similar coupling arguments. □

We now present some remarks on Theorem 3.1. The theorem shows that each of the four components of the $i\mathcal{S}$-graph experiences a phase transition at some critical density $\lambda_c^\diamond$. These critical densities are *nontrivial*, in the sense that $0 < \lambda_c^\diamond < \infty$. As we increase the density $\lambda_\ell$ of legitimate nodes, the component $\mathcal{K}^{\text{weak}}(0)$ percolates first, then $\mathcal{K}^{\text{out}}(0)$ or $\mathcal{K}^{\text{in}}(0)$, and finally $\mathcal{K}^{\text{strong}}(0)$. Furthermore, percolation can occur for any prescribed infimum secrecy rate $\varrho$, as long as it is below the channel capacity of a link with zero length, i.e., $\varrho_{\max}$. This has different implications depending on the type of path loss model, as presented in Section II-A:

- If $g(0) = \infty$, percolation can occur for any arbitrarily large secrecy requirement $\varrho$, as long as the density $\lambda_\ell$ of legitimate nodes is made large enough.
- If $g(0) < \infty$, percolation cannot occur if the threshold $\varrho$ is set above $\varrho_{\max} = \log_2(1 + \mathsf{SNR} \cdot g(0))$, where $\mathsf{SNR} \triangleq \frac{P}{\sigma^2}$. To ensure percolation for such $\varrho$, the signal-to-noise-ratio $\mathsf{SNR}$ must be

1414

increased until $\varrho_{\max}(\mathsf{SNR})$ decreases below the desired $\varrho$. At that point, the density $\lambda_\ell$ can then be increased to achieve percolation.

Note that the theorem holds for any channel gain function $g(r)$ satisfying Conditions 1–3 in Section II-A, including bounded and unbounded models. Concerning the density $\lambda_\mathrm{e}$ of eavesdroppers, the theorem says that as long as $\varrho < \varrho_{\max}$, percolation can occur even in scenarios with arbitrarily dense eavesdroppers. This can be achieved just by deploying more legitimate nodes, so that $\lambda_\ell$ is large enough.

Operationally, the theorem is important because it shows that long-range secure communication over multiple hops is still feasible, even in the presence of arbitrarily dense eavesdroppers. In particular, if we limit communication to the secure links whose MSR exceeds $\varrho$ (chosen such that $\varrho < \varrho_{\max}$), then for $\lambda_\ell$ large enough, a component with an infinite number of securely-connected nodes arises. Those nodes are able to communicate with strong secrecy (in the sense of Definition 2.1), at a rate greater than $\varrho$ bits per complex channel use. The specific type of the secure connection (e.g., unidirectional or bidirectional) depends on the graph component under consideration: out, in, weak, or strong component.

## C. Proof of Lemma 3.1

In this section, we introduce a few definitions and propositions, which are then used to prove the lemma. Note that the graph $G^{\mathrm{weak}}(\varrho)$ depends on $\varrho$, and it is sufficient to show that $G^{\mathrm{weak}}(\varrho)$ *for the case of $\varrho = 0$ does not percolate for sufficiently small $\lambda_\ell$*. This is because for larger $\varrho$ the connectivity of $G^{\mathrm{weak}}(\varrho)$ is worse and thus $G^{\mathrm{weak}}(\varrho)$ certainly does not percolate either. We then proceed in two intermediate steps. First, we map the continuous $i\mathcal{S}$-graph $G$ onto a discrete hexagonal lattice $\mathcal{L}_\mathrm{h}$, and declare a face in $\mathcal{L}_\mathrm{h}$ to be closed in such a way that the absence of face percolation in $\mathcal{L}_\mathrm{h}$ implies the absence of continuum percolation in $G^{\mathrm{weak}}$. Second, we show that discrete face percolation does not occur in $\mathcal{L}_\mathrm{h}$ for sufficiently small (but nonzero) $\lambda_\ell$. The details are presented next.

*1) Mapping on a Lattice:* We start with some definitions. Let $\mathcal{L}_\mathrm{h}$ be an hexagonal lattice as depicted in Fig. 3, where each face is a regular hexagon with side length $\delta$. Each face has a *state*, which can be either *open* or *closed*. A set of faces (e.g., a path or a circuit) in $\mathcal{L}_\mathrm{h}$ is said to be open iff all the faces that form the set are open. We now define when a face is *closed* based on how the processes $\Pi_\ell$ and $\Pi_\mathrm{e}$ behave inside that face.

*Definition 3.1 (Closed Face in $\mathcal{L}_h$):* A face $\mathcal{H}$ in $\mathcal{L}_h$ is said to be *closed* iff all the following conditions are met:

1) Each of the 6 equilateral triangles $\{\mathcal{T}_i\}_{i=1}^{6}$ that compose the hexagon $\mathcal{H}$ has at least one eavesdropper.
2) The hexagon $\mathcal{H}$ is free of legitimate nodes.

The above definition was chosen such that discrete face percolation in $\mathcal{L}_h$ can be tied to continuum percolation in $G^{\text{weak}}$, as given by the following proposition.

*Proposition 3.1 (Circuit Coupling):* If there exists a closed circuit in $\mathcal{L}_h$ surrounding the origin, then the component $\mathcal{K}^{\text{weak}}(0)$ is finite.

*Proof:* Assume there is a closed circuit $\mathcal{C}$ in $\mathcal{L}_h$ surrounding the origin, as depicted in Fig. 4. This implies that the open component in $\mathcal{L}_h$ containing 0, denoted by $\mathcal{K}^{\mathcal{L}_h}(0)$, must be finite. Since the area of the region $\mathcal{K}^{\mathcal{L}_h}(0)$ is finite, the continuous graph $G^{\text{weak}}$ has a finite number of vertices falling inside this region. Thus, to prove that $\mathcal{K}^{\text{weak}}(0)$ is finite, we just need to show that no edges of $G^{\text{weak}}$ cross the circuit $\mathcal{C}$. Consider Fig. 3, and suppose that the shaded faces are part of the closed circuit $\mathcal{C}$. Let $x_1, x_2$ be two legitimate nodes such that $x_1$ falls on an open face on the inner side of $\mathcal{C}$, while $x_2$ falls on the outer side of $\mathcal{C}$ (note that Definition 3.1 specifies that the closed faces in $\mathcal{C}$ cannot contain legitimate nodes). Clearly, the most favorable situation for $x_1, x_2$ being able to establish an edge across $\mathcal{C}$ is the one depicted in Fig. 3. The edge $\overline{x_1 x_2}$ exists in $G^{\text{weak}}$ iff either $\mathcal{B}_{x_1}(\delta)$ or $\mathcal{B}_{x_2}(\delta)$ are free of eavesdroppers.[12] This condition does not hold, since Definition 3.1 guarantees that at least one eavesdropper is located inside the triangle $\mathcal{T}_i \subset \mathcal{B}_{x_1}(\delta) \cap \mathcal{B}_{x_2}(\delta)$. Thus, no edges of $G^{\text{weak}}$ cross the circuit $\mathcal{C}$, which implies that $\mathcal{K}^{\text{weak}}(0)$ has finite size. □

*2) Discrete Percolation:* Having performed an appropriate mapping from a continuous to a discrete model, we now analyze discrete face percolation in $\mathcal{L}_h$.

*Proposition 3.2 (Discrete Percolation in $\mathcal{L}_h$):* If the parameters $\lambda_\ell, \lambda_e, \delta$ satisfy

$$\left(1 - e^{-\lambda_e \frac{\sqrt{3}}{4}\delta^2}\right)^6 \cdot e^{-\lambda_\ell \frac{3\sqrt{3}}{2}\delta^2} > \frac{1}{2}, \tag{20}$$

then the origin is a.s. surrounded by a closed circuit in $\mathcal{L}_h$.

*Proof:* According to Definition 3.1, the state of a face $\mathcal{H}$ in $\mathcal{L}_h$ does not depend on the behaviour of the processes $\Pi_\ell$ and $\Pi_e$ outside $\mathcal{H}$. Because the processes are Poisson, the state of different faces is then independent. Then, the model introduced in Section III-C1 can be seen as a face percolation model on the hexagonal lattice $\mathcal{L}_h$, where each face is closed independently of other faces with probability

$$p \triangleq \mathbb{P}\{\text{face } \mathcal{H} \text{ of } \mathcal{L}_h \text{ is closed}\}$$
$$= \mathbb{P}\left\{\left(\bigwedge_{i=1}^{6} \Pi_e\{\mathcal{T}_i\} \geq 1\right) \wedge \Pi_\ell\{\mathcal{H}\} = 0\right\}$$
$$= \left(1 - e^{-\lambda_e \frac{\sqrt{3}}{4}\delta^2}\right)^6 \cdot e^{-\lambda_\ell \frac{3\sqrt{3}}{2}\delta^2}, \tag{21}$$

where we used the independence between the processes $\Pi_\ell$ and $\Pi_e$, and the fact that $\mathbb{A}\{\mathcal{T}_i\} = \frac{\sqrt{3}}{4}\delta^2$ and $\mathbb{A}\{\mathcal{H}\} = \frac{3\sqrt{3}}{2}\delta^2$. Face percolation on the hexagonal lattice can be equivalently described as site percolation on the triangular lattice. In particular, recall that if

$$\mathbb{P}\{\mathcal{H} \text{ is open}\} < \frac{1}{2}, \tag{22}$$

then the *open* component in $\mathcal{L}_h$ containing the origin is a.s. finite [50, Ch. 5, Thm. 8], and so the origin is a.s. surrounded by a *closed* circuit in $\mathcal{L}_h$. Combining (21) and (22), we obtain (20). □

---

[12]We use $\mathcal{B}_x(\rho) \triangleq \{y \in \mathbb{R}^2 : |y - x| \leq \rho\}$ to denote the closed two-dimensional ball centered at point $x$, with radius $\rho$.

We now use the propositions to finalize the proof of Lemma 3.1, whereby $p_\infty^{\text{weak}}(\lambda_\ell) = 0$ for sufficiently small (but nonzero) $\lambda_\ell$.

*Proof of Lemma 3.1:* For any fixed $\lambda_e$, it is easy to see that condition (20) can always be met by making the hexagon side $\delta$ large enough, and the density $\lambda_\ell$ small enough (but nonzero). For any such choice of parameters $\lambda_\ell, \lambda_e, \delta$ satisfying (20), the origin is a.s. surrounded by a closed circuit in $\mathcal{L}_h$ (by Proposition 3.2), and the component $\mathcal{K}^{\text{weak}}(0)$ is a.s. finite (by Proposition 3.1), i.e., $p_\infty^{\text{weak}}(\lambda_\ell) = 0$. □

### D. Proof of Lemma 3.2

In this section, we introduce a few definitions and propositions, which are then used to prove the lemma. We proceed in two intermediate steps. First, we associate with our continuous $i\mathcal{S}$-graph $G$ a discrete square lattice $\mathcal{L}_s$ as well as its dual $\mathcal{L}'_s$, and declare an edge in $\mathcal{L}_s$ to be open in such a way that discrete edge percolation in $\mathcal{L}'_s$ implies continuum percolation in $G^{\text{strong}}$. Second, we show that discrete edge percolation occurs in $\mathcal{L}'_s$ for sufficiently large (but finite) $\lambda_\ell$. The details are presented next.

*1) Mapping on a Lattice:* We start with some definitions. Let $\mathcal{L}_{\mathrm{s}} \triangleq d \cdot \mathbb{Z}^2$ be a square lattice with edge length $d$. Let $\mathcal{L}'_{\mathrm{s}}$ be the dual lattice of $\mathcal{L}_{\mathrm{s}}$, constructed by placing a vertex in the center of every square of $\mathcal{L}_{\mathrm{s}}$, and placing an edge $a'$ across every edge $a$ of $\mathcal{L}_{\mathrm{s}}$. Since $\mathcal{L}_{\mathrm{s}}$ is a square lattice, it is clear that $\mathcal{L}'_{\mathrm{s}} = \mathcal{L}_{\mathrm{s}} + \left(\frac{d}{2}, \frac{d}{2}\right)$, as depicted in Fig. 5. We make the origin of the coordinate system coincide with a vertex of $\mathcal{L}'_{\mathrm{s}}$. Each edge has a *state*, which can be either *open* or *closed*. We declare an edge $a'$ of $\mathcal{L}'_{\mathrm{s}}$ to be open iff its dual edge $a$ in $\mathcal{L}_{\mathrm{s}}$ is open. Furthermore, a set of edges (e.g., a path or a circuit) in $\mathcal{L}_{\mathrm{s}}$ or $\mathcal{L}'_{\mathrm{s}}$ is said to be open iff all the edges that form the set are open.

We now specify when an edge of $\mathcal{L}_{\mathrm{s}}$ (and therefore of $\mathcal{L}'_{\mathrm{s}}$) is *open* based on how the processes $\Pi_\ell$ and $\Pi_{\mathrm{e}}$ behave in the neighborhood of that edge. Consider Fig. 6, where $a$ denotes an edge in $\mathcal{L}_{\mathrm{s}}$, and $\mathcal{S}_1(a)$ and $\mathcal{S}_2(a)$ denote the two squares adjacent to $a$. Let $\{v_i\}_{i=1}^4$ denote the four vertices of the rectangle $\mathcal{S}_1(a) \cup \mathcal{S}_2(a)$. We then have the following definition.

*Definition 3.2 (Open Edge in $\mathcal{L}_{\mathrm{s}}$):* An edge $a$ in $\mathcal{L}_{\mathrm{s}}$ is said to be *open* iff all the following conditions are met:

1) Each square $\mathcal{S}_1(a)$ and $\mathcal{S}_2(a)$ adjacent to $a$ has at least one legitimate node.
2) The region $\mathcal{Z}(a)$ is free of eavesdroppers, where $\mathcal{Z}(a)$ is smallest rectangle such that $\bigcup_{i=1}^4 \mathcal{B}_{v_i}(r_{\mathrm{free}}) \subset \mathcal{Z}(a)$ with[13]

$$r_{\mathrm{free}} \triangleq g^{-1}\left(2^{-\varrho} g(\sqrt{5}d) - \frac{\sigma^2}{P}(1 - 2^{-\varrho})\right). \tag{23}$$

The above definition was chosen such that discrete edge percolation in $\mathcal{L}'_{\mathrm{s}}$ can be tied to continuum percolation in $G^{\mathrm{strong}}$, as given by the following two propositions.

*Proposition 3.3 (Two-Square Coupling):* If $a$ is an open edge in $\mathcal{L}_{\mathrm{s}}$, then all legitimate nodes inside $\mathcal{S}_1(a) \cup \mathcal{S}_2(a)$ form a single connected component in $G^{\mathrm{strong}}$.

*Proof:* If two legitimate nodes $x_1, x_2$ are to be placed inside the region $\mathcal{S}_1(a) \cup \mathcal{S}_2(a)$, the most unfavorable configuration in terms of MSR is the one where the distance $|x_1 - x_2|$ is maximized, i.e., $x_1, x_2$ are on opposite corners of the rectangle $\mathcal{S}_1(a) \cup \mathcal{S}_2(a)$ and thus $|x_1 - x_2| = \sqrt{5}d$. In such configuration, we see from (7) that the edge $\overrightarrow{x_1 x_2}$ exists in $G$ iff $g(|x_i - x_j|) > 2^\varrho g(|x_i - e^*|) + \frac{\sigma^2}{P}(2^\varrho - 1)$, where $e^*$ is the eavesdropper closest to $x_1$. This is equivalent to

---

[13]To ensure that $r_{\mathrm{free}}$ in (23) is well-defined, in the rest of the paper we assume that $d$ is chosen such that $d < \frac{1}{\sqrt{5}} g^{-1}\left(\frac{\sigma^2}{P}(2^\varrho - 1)\right)$.

requiring that

$$|x_1 - e^*| > g^{-1}\left(2^{-\varrho}g(\sqrt{5}d) - \frac{\sigma^2}{P}(1 - 2^{-\varrho})\right)$$

$$\triangleq r_{\text{free}},$$

which is a well-defined quantity if $d$ is chosen such that $d < \frac{1}{\sqrt{5}}g^{-1}\left(\frac{\sigma^2}{P}(2^{\varrho} - 1)\right)$. As a result, the edge $\overline{x_1 x_2}$ exists in $G^{\text{strong}}$ iff both $\mathcal{B}_{x_1}(r_{\text{free}})$ and $\mathcal{B}_{x_2}(r_{\text{free}})$ are free of eavesdroppers. Now, if $\mathcal{Z}(a)$ is the smallest rectangle containing the region $\bigcup_{i=1}^{4}\mathcal{B}_{v_i}(r_{\text{free}})$, where $v_i$ are the vertices of $\mathcal{S}_1(a) \cup \mathcal{S}_2(a)$, then the condition $\Pi_e\{\mathcal{Z}(a)\} = 0$ ensures the edge $\overline{x_i x_j}$ exists in $G^{\text{strong}}$ for any $x_i, x_j \in \mathcal{S}_1(a) \cup \mathcal{S}_2(a)$, and thus all legitimate nodes inside $\mathcal{S}_1(a) \cup \mathcal{S}_2(a)$ form a single connected component in $G^{\text{strong}}$. $\square$

*Proposition 3.4 (Component Coupling):* If the open component in $\mathcal{L}'_s$ containing the origin is infinite, then the component $\mathcal{K}^{\text{strong}}(0)$ is also infinite.

*Proof:* Consider Fig. 7. Let $\mathcal{P} = \{a'_i\}$ denote a path of open edges $\{a'_i\}$ in $\mathcal{L}'_s$. By the definition of dual lattice, the path $\mathcal{P}$ uniquely defines a sequence $\mathcal{S} = \{\mathcal{S}_i\}$ of *adjacent* squares in $\mathcal{L}_s$, separated by open edges $\{a_i\}$ (the duals of $\{a'_i\}$). Then, each square in $\mathcal{S}$ has at least one legitimate node (by Definition 3.2), and all legitimate nodes falling inside the region associated with $\mathcal{S}$ form a single connected component in $G^{\text{strong}}$ (by Proposition 3.3). Now, let $\mathcal{K}^{\mathcal{L}'_s}(0)$ denote the open component in $\mathcal{L}'_s$ containing $0$, i.e., the set of vertexes in $\mathcal{L}'_s$ that are connected to $0$ by some path. Because of the argument just presented, we have $|\mathcal{K}^{\mathcal{L}'_s}(0)| \leq |\mathcal{K}^{\text{strong}}(0)|$. Thus, if $|\mathcal{K}^{\mathcal{L}'_s}(0)| = \infty$, then $|\mathcal{K}^{\text{strong}}(0)| = \infty$. $\square$

*2) Discrete Percolation:* Having performed an appropriate mapping from a continuous to a discrete model, we now analyze discrete edge percolation in $\mathcal{L}'_s$. Let $N_s$ be the number of squares that compose the rectangle $\mathcal{Z}(a)$ introduced in Definition 3.2. We first study the behavior of paths in $\mathcal{L}_s$ with the following proposition.

*Proposition 3.5 (Geometric Bound):* The probability that a given path of $\mathcal{L}_s$ with length $n$ is *closed* is bounded by

$$\mathbb{P}\{\text{path of } \mathcal{L}_s \text{ with length } n \text{ is closed}\} \leq q^{n/N_e}, \tag{24}$$

where $N_e$ is a finite integer and

$$q = 1 - (1 - e^{-\lambda_\ell d^2})^2 \cdot e^{-\lambda_e N_s d^2} \tag{25}$$

is the probability that an edge of $\mathcal{L}_\text{s}$ is closed.

*Proof:* Using Definition 3.2, we can write

$$q \triangleq \mathbb{P}\{\text{edge } a \text{ of } \mathcal{L}_\text{s} \text{ is closed}\}$$
$$= 1 - \mathbb{P}\{\Pi_\ell\{\mathcal{S}_1(a)\} \geq 1 \wedge \Pi_\ell\{\mathcal{S}_2(a)\} \geq 1 \wedge \Pi_\text{e}\{\mathcal{Z}(a)\} = 0\}$$
$$= 1 - (1 - e^{-\lambda_\ell d^2})^2 \cdot e^{-\lambda_\text{e} N_\text{s} d^2},$$

where we used the properties of the independent processes $\Pi_\ell$ and $\Pi_\text{e}$. This is the result in (25). Now, letting $\mathcal{P} = \{a_i\}_{i=1}^n$ denote a path of $\mathcal{L}_\text{s}$ with length $n$ and edges $\{a_i\}$, we wish to obtain an upper bound on $\mathbb{P}\{\mathcal{P} \text{ is closed}\}$. Considering two edges $a_i, a_j \in \mathcal{P}$, the states of these edges are statistically independent iff

$$\mathcal{Z}(a_i) \cap \mathcal{Z}(a_j) = \emptyset. \tag{26}$$

We consider a subset $\mathcal{Q}$ of edges in $\mathcal{P}$, constructed in the following way. Start with the first edge $a_1 \in \mathcal{P}$, whose associated region is $\mathcal{Z}(a_1)$, and add it to the subset $\mathcal{Q}$. Now, determine the next edge $a_k \in \mathcal{P}$ such that $\mathcal{Z}(a_1) \cap \mathcal{Z}(a_k) = \emptyset$, and add it to the subset $\mathcal{Q}$. Repeat the process until there are no more edges in path $\mathcal{P}$. By construction, it is easy to see that $\mathcal{Q} \subseteq \mathcal{P}$, and any two edges in $\mathcal{Q}$ have independent states since they satisfy (26). Thus,

$$\mathbb{P}\{\mathcal{P} \text{ is closed}\} \leq \mathbb{P}\{\mathcal{Q} \text{ is closed}\}$$
$$= q^m,$$

where $m = \#\mathcal{Q}$. After careful analysis of Fig. 6, we observe that the rectangle $\mathcal{Z}(a)$ has dimensions $M \times (M+1)$ squares, where $M = 2\left\lceil \frac{r_\text{free}}{d} \right\rceil + 1$. Furthermore, starting in edge $a$, we can count at most $N_\text{e} = 8M^2 - 1$ edges (including $a$ itself) until we reach the next element of $\mathcal{Q}$. As a result,

$$m \geq \left\lceil \frac{n}{N_\text{e}} \right\rceil \geq \frac{n}{N_\text{e}},$$

and the desired upper bound becomes

$$\mathbb{P}\{\mathcal{P} \text{ is closed}\} \leq q^{n/N_\text{e}},$$

which is the result in (24). Since $r_\text{free}$ in (23) is guaranteed to be finite, then $N_\text{s}$ and $N_\text{e}$ are also finite (although possibly large). □

We have just shown that, although there is dependence between the state of different edges of $\mathcal{L}_\mathrm{s}$, the probability of a path of length $n$ being closed decays geometrically as $q^{n/N_\mathrm{e}}$. We can now use a Peierls argument to study the existence of an infinite component.[14]

*Proposition 3.6 (Discrete Percolation in $\mathcal{L}'_\mathrm{s}$):* If the probability $q$ satisfies

$$q < \left(\frac{11 - 2\sqrt{10}}{27}\right)^{N_\mathrm{e}}, \tag{27}$$

then

$$\mathbb{P}\{\text{open component in } \mathcal{L}'_\mathrm{s} \text{ containing } 0 \text{ is infinite}\} > 0. \tag{28}$$

*Proof:* We start with the key observation that the open component in $\mathcal{L}'_\mathrm{s}$ containing 0 is *finite* iff there is a closed circuit in $\mathcal{L}_\mathrm{s}$ surrounding 0. This is best seen by inspecting Fig. 7, where the origin is surrounded by a necklace of closed edges in $\mathcal{L}'_\mathrm{s}$, which block all possible routes in $\mathcal{L}_\mathrm{s}$ from the origin to infinity. Thus, the inequality in (28) is equivalent to $\mathbb{P}\{\exists \text{ closed circuit in } \mathcal{L}_\mathrm{s} \text{ surrounding } 0\} < 1$. Let $\rho(n)$ denote the possible number of circuits of length $n$ in $\mathcal{L}_\mathrm{s}$ surrounding 0 (a deterministic quantity). Let $\kappa(n)$ denote the number of *closed* circuits of length $n$ in $\mathcal{L}_\mathrm{s}$ surrounding 0 (a random variable). From combinatorial arguments, it can be shown [52, (1.17)] that

$$\rho(n) \leq 4n3^{n-2}.$$

Then, for a fixed $n$,

$$\mathbb{P}\{\kappa(n) \geq 1\} \leq \rho(n)\mathbb{P}\{\text{path of } \mathcal{L}_\mathrm{s} \text{ with length } n \text{ is closed}\}$$
$$\leq 4n3^{n-2}q^{n/N_\mathrm{e}},$$

---

[14] A "Peierls argument", so-named in honour of Rudolf Peierls and his 1936 article on the Ising model [51], refers to an approach based on enumeration. For a simple example, see [52, pp. 16–19].

where we used the union bound and Proposition 3.5. Also,

$$\mathbb{P}\{\exists \text{ closed circuit in } \mathcal{L}_\text{s} \text{ surrounding } 0\} = \mathbb{P}\{\kappa(n) \geq 1 \text{ for some } n\}$$

$$\leq \sum_{n=1}^{\infty} \mathbb{P}\{\kappa(n) \geq 1\}$$

$$\leq \sum_{n=1}^{\infty} 4n 3^{n-2} q^{n/N_\text{e}}$$

$$= \frac{4 q^{1/N_\text{e}}}{3(1 - 3 q^{1/N_\text{e}})^2}, \quad (29)$$

for $q < \left(\frac{1}{3}\right)^{N_\text{e}}$. We see that if $q$ satisfies (27), then (29) is strictly less than one, and (28) follows. □

We now use the propositions to finalize the proof of Lemma 3.2, whereby $p_\infty^{\text{strong}}(\lambda_\ell) > 0$ for sufficiently large (but finite) $\lambda_\ell$.

*Proof of Lemma 3.2:* For any fixed $\lambda_\text{e}$, it is easy to see the probability $q$ in (25) can be made small enough to satisfy condition (27), by making the edge length $d$ sufficiently small, and the density $\lambda_\ell$ sufficiently large (but finite). For any such choice of parameters $\lambda_\ell, \lambda_\text{e}, d$ satisfying (27), the open component in $\mathcal{L}'_\text{s}$ containing $0$ is infinite with positive probability (by Proposition 3.6), and the component $\mathcal{K}^{\text{strong}}(0)$ is also infinite with positive probability (by Proposition 3.4), i.e., $p_\infty^{\text{strong}}(\lambda_\ell) > 0$. □

*E. Simulation Results*

In this section, we obtain additional insights about percolation in the $i\mathcal{S}$-graph via Monte Carlo simulation. Specifically, we aim to evaluate the percolation probabilities $p_\infty^\diamond$ as a function of the density $\lambda_\ell$ of legitimate nodes, and thus estimate the corresponding critical densities $\lambda_\text{c}^\diamond$.

We now describe the simulation procedure for evaluating the percolation probabilities. We consider a square $\mathcal{R}$ with dimensions $\sqrt{A} \times \sqrt{A}$. The area $A$ is adjusted according to $A = \frac{N_\ell}{\lambda_\ell}$, where the average number $N_\ell$ of legitimate nodes in $\mathcal{R}$ is kept fixed. This ensures that the simulation time is approximately constant with respect to the parameter $\lambda_\ell$. In the simulations, we use $N_\ell = 5000$ nodes and $\lambda_\text{e} = 1\,\text{m}^{-2}$. We first place $\Pi_\ell\{\mathcal{R}\} \sim \mathcal{P}(\lambda_\ell A)$ legitimate nodes and $\Pi_\text{e}\{\mathcal{R}\} \sim \mathcal{P}(\lambda_\text{e} A)$ legitimate nodes inside $\mathcal{R}$, uniformly and independently.[15] The $i\mathcal{S}$-graph $G =$

---

[15] We use $\mathcal{P}(\mu)$ to denote a discrete Poisson distribution with mean $\mu$.

$\{\Pi_\ell, \mathcal{E}\}$ is then established using as edge set

$$\mathcal{E} = \left\{\overrightarrow{x_i x_j} : g(d(x_i, x_j)) > 2^\varrho g(d(x_i, e^*)) + \frac{\sigma^2}{P}(2^\varrho - 1), \quad e^* = \operatorname*{argmin}_{e_k \in \Pi_e} d(x_i, e_k)\right\}, \qquad (30)$$

where $d(\cdot, \cdot)$ is a toroidal distance metric [53], [54].[16] After the $i\mathcal{S}$-graph is established, we determine the various components in $G$, $G^{\text{weak}}$, and $G^{\text{strong}}$. The percolation probabilities are then calculated using the result in Appendix B:

$$p_\infty^\diamond = \frac{\mathbb{E}\{N_\infty^\diamond\}}{\lambda_\ell A} \approx \frac{\mathbb{E}\{N_{\text{largest}}^\diamond\}}{\lambda_\ell A}, \qquad (31)$$

where $\diamond$ refers to the weak or strong component, and $N_{\text{largest}}^\diamond$ is the size of the largest component of the weak or strong $i\mathcal{S}$-graph restricted to the region $\mathcal{R}$. The need for the approximation is the following: since the simulation region $\mathcal{R}$ is finite, it is not possible to determine whether a node $x$ in $\mathcal{R}$ has an infinite component $\mathcal{K}^\diamond(x)$ or not. Thus, the number of nodes in $\mathcal{R}$ whose component $\mathcal{K}^\diamond(x)$ is infinite is approximated by the number of nodes belonging to the largest component inside $\mathcal{R}$, similarly to [34]. A little reflection also shows that the above approximation is only reasonable for the weak and strong components, but not for the out- and in-components, and so we consider only the first two. The expectation in (31) is computed over an ensemble of $20$ spatial realizations of $\Pi_\ell$ and $\Pi_e$.

Figure 8 shows the simulated percolation probabilities for the weak and strong components of the $i\mathcal{S}$-graph, versus the density $\lambda_\ell$ of legitimate nodes. It considers the simplest case of $\varrho = 0$, for which the percolation probabilities depend only on the ratio $\frac{\lambda_\ell}{\lambda_e}$.[17] As predicted by Theorem 3.1, the weak and strong components experience phase transitions as $\lambda_\ell$ is increased. Indeed, the curves $p_\infty^\diamond(\lambda_\ell)$ exhibit a fast increase immediately after the critical density $\lambda_c^\diamond$ is reached. The reason why $p_\infty^\diamond(\lambda_\ell)$ is not exactly zero for $\lambda_\ell < \lambda_c^\diamond$ is the approximation made in (31): even though there is no infinite component in such regime, there is a nonzero probability that large finite components arise, and these contribute to a nonzero $\mathbb{E}\{N_{\text{largest}}^\diamond\}$. Figure 8 suggests that $\lambda_c^{\text{weak}} \approx 3.4\,\text{m}^{-2}$ and $\lambda_c^{\text{strong}} \approx 6.2\,\text{m}^{-2}$, for the case of $\lambda_e = 1\,\text{m}^{-2}$ and $\varrho = 0$. Operationally, this means that if long-range bidirectional secure communication is desired in a wireless network, the density of legitimate nodes must be at least $6.2$ times that of the eavesdroppers. In practice, this ratio must be even larger, because a security requirement greater than $\varrho = 0$ is typically required.[18] Furthermore, increasing $\lambda_\ell$ also leads to an increased average fraction of nodes $p_\infty^{\text{strong}}$ which belong to the infinite component, thus ensuring better connectivity of the network.

---

[16]The use of the Euclidean metric $|x_i - x_j|$ over the finite region $\mathcal{R}$ would give rise to boundary effects, since legitimate nodes near the borders would be isolated with higher probability than the nodes in the middle. The toroidal distance metric, on the other hand, transforms the square region $\mathcal{R}$ into a torus, and minimizes such boundary effects in the simulations. Other edge correction methods are discussed in [54].

[17]The proof of this fact is entirely analogous to the proof of [31, Property 3.1].

Figure 9 illustrates the subcritical and supercritical phases of the $i\mathcal{S}$-graph. In Fig. 9(a), we have $\frac{\lambda_\ell}{\lambda_e} = 2$, and the $i\mathcal{S}$-graph exhibits only small, bounded clusters of legitimate nodes. This is in agreement with Fig. 8, which suggests that for a ratio of $\frac{\lambda_\ell}{\lambda_e} = 2$, all four out, in, weak, and strong components are subcritical. In Fig. 9(b), we have $\frac{\lambda_\ell}{\lambda_e} = 10$, and the $i\mathcal{S}$-graph exhibits a large cluster of connected nodes. This also agrees with Fig. 8, which suggests that for a ratio of $\frac{\lambda_\ell}{\lambda_e} = 10$, all four out, in, weak, and strong components are supercritical.

Figure 10 illustrates the dependence of the percolation probability $p_\infty^{\text{weak}}$ on the secrecy rate threshold $\varrho$. As expected, we observe that the critical density $\lambda_c^{\text{weak}}$ is increasing with respect to $\varrho$. This is because as we increase the threshold $\varrho$, the requirement $\mathcal{R}_\mathsf{s}(x_i, x_j) > \varrho$ for any two nodes $x_i, x_j$ to be securely connected becomes stricter. Thus, the connectivity of the $i\mathcal{S}$-graph becomes worse and a higher density of legitimate nodes is needed for percolation.

Figure 11 illustrates the dependence of the percolation probability $p_\infty^{\text{weak}}$ on the wireless propagation effects, such as lognormal shadowing and Rayleigh fading. From the curves, we observe that $\lambda_c^{\text{weak}}(\text{lognormal}) < \lambda_c^{\text{weak}}(\text{Rayleigh}) < \lambda_c^{\text{weak}}(\text{deterministic})$, i.e., the randomness of the wireless channel—as observed in realistic environments—*improves* long-range secure connectivity, by decreasing the critical density at which percolation occurs. This phenomenon contrasts with the behavior of local connectivity, where channel randomness *does not* change the PMF of the out-degree $N_{\text{out}}$ [31]. However, channel randomness *does* affect the PMF of the in-degree $N_{\text{in}}$, as well as the statistical dependencies between the degrees of different nodes, and therefore affects the properties of multi-hop connectivity.[19] Furthermore, we conclude that by assuming the absence of fading—as we do in the majority of this chapter to ensure mathematical tractability—we are in effect considering the *most pessimistic scenario* in terms of long-range secure connectivity.

---

[18]The critical densities $\lambda_c^\diamond(\lambda_e, \varrho)$ are non-decreasing functions of $\lambda_e$ and $\varrho$, as can be shown using a coupling argument similar to the proof of Lemma 3.3.

[19]Note that in the absence of fading, the degrees of different legitimate nodes are *statistically dependent*, because different edges depend on a *common* underlying process $\Pi_e$ of eavesdroppers. For example, given that a legitimate node is isolated (due to the proximity of an eavesdropper), then it is also likely that nearby legitimate nodes will also be isolated. By introducing random fading, such dependence on the underlying eavesdropper process is decreased, and multi-hop connectivity is improved.

## IV. Full Connectivity in the Poisson $i\mathcal{S}$-Graph

In the previous sections, we studied percolation in the $i\mathcal{S}$-graph defined over the infinite plane. We showed that for some combinations of the parameters $(\lambda_\ell, \lambda_e, \varrho)$, the regime is supercritical and an infinite component arises. However, the existence of an infinite component does not ensure connectivity between any two nodes, e.g., one node inside the infinite component cannot communicate with a node outside. In this sense, percolation ensures only *partial connectivity* of the network. In some scenarios, it is of interest to guarantee *full connectivity*, i.e., that all nodes can communicate with each other, possibly through multiple hops. Note, however, that for networks defined over an infinite region, the probability of full connectivity is exactly zero. Thus, to study of full connectivity, we need to restrict our attention to a finite region $\mathcal{R}$.

Throughout this section, we consider the simplest case of $\varrho = 0$, i.e., the *existence* of secure links with a positive (but possibly small) MSR. Because this scenario is characterized by the simple geometric description in (8), it provides various insights that are useful in understanding more complex scenarios.[20] Furthermore, the case of $\varrho = 0$ represents the *most favorable scenario* in terms of full connectivity, since a higher security requirement $\varrho$ leads to degraded connectivity.

In what follows, we start by defining full connectivity in the $i\mathcal{S}$-graph. We then characterize the exact asymptotic behavior of full connectivity in the limit of a large density of legitimate nodes. Lastly, we derive simple, explicit expressions that closely approximate the probability of full in- and out-connectivity for a finite density of legitimate nodes, and determine the accuracy of such approximations using simulations.

### A. Definitions

Since the $i\mathcal{S}$-graph is a directed graph, we start by distinguishing between full out- and in-connectivity with the following definitions.

---

[20]Specifically, the case of $\varrho = 0$ brings the following mathematical simplifications. First, the $i\mathcal{S}$-graph is completely independent of channel gain function $g(r)$, thus no assumptions about the propagation model are needed. Second, there exist simple (often closed-form) expressions for characterizing local connectivity [31] which will be useful in analyzing full connectivity.

*Definition 4.1 (Full Out-Connectivity):* A legitimate node $x_i \in \Pi_\ell \cap \mathcal{R}$ is *fully out-connected* with respect to a region $\mathcal{R}$ if in the $i\mathcal{S}$-graph $G = \{\Pi_\ell, \mathcal{E}\}$ there exists a directed path from $x_i$ to *every* node $x_j \in \Pi_\ell \cap \mathcal{R}$, for $x_j \neq x_i$.

*Definition 4.2 (Full In-Connectivity):* A legitimate node $x_i \in \Pi_\ell \cap \mathcal{R}$ is *fully in-connected* with respect to a region $\mathcal{R}$ if in the $i\mathcal{S}$-graph $G = \{\Pi_\ell, \mathcal{E}\}$ there exists a directed path to $x_i$ from *every* node $x_j \in \Pi_\ell \cap \mathcal{R}$, for $x_j \neq x_i$.[21]

Since the $i\mathcal{S}$-graph is a random graph, we can consider the probabilities of a node $x_i$ being fully out- or in-connected. For analysis purposes, we consider that probe legitimate node (node 0) placed at the origin of the coordinate system, i.e., $x_{\text{probe}} = 0 \in \mathcal{R}$. We then define $p_{\text{out-con}}$ and $p_{\text{in-con}}$ as the probability that node 0 is, respectively, fully out- and fully in-connected. These probabilities are a deterministic function of the densities $\lambda_\ell$ and $\lambda_\text{e}$, and the area $A$ of region $\mathcal{R}$. Our goal is to characterize $p_{\text{out-con}}$ and $p_{\text{in-con}}$.

## B. Full Connectivity: Asymptotic Regime

In this section, we focus on the asymptotic behavior of secure connectivity as we increase the density of legitimate nodes. Specifically, for a fixed region of area $A$ and a fixed density $\lambda_\text{e}$ of eavesdroppers, we would like to determine if by increasing $\lambda_\ell \to \infty$, we can asymptotically achieve full in- and out-connectivity with probability one.[22] Note that a.s. full connectivity can only be achieved *asymptotically*, since for any finite $\lambda_\ell$, the probabilities $p_{\text{out-con}}$ and $p_{\text{in-con}}$ are strictly less than one.

*Definition 4.3 (Asymptotic Out-Connectivity):* A legitimate node $x \in \Pi_\ell \cap \mathcal{R}$ is *asymptotically out-connected* with respect to a region $\mathcal{R}$ with area $A$ if $\lim_{\lambda_\ell \to \infty} p_{\text{out-con}} = 1$, for any $\lambda_\text{e} > 0$ and $A > 0$.

---

[21]Note that these two definitions imply that that legitimate nodes *outside* the region $\mathcal{R}$ can act as relays between legitimate nodes *inside* $\mathcal{R}$. Essentially, we are considering the $i\mathcal{S}$-graph defined on the infinite plane, but are only interested in the full connectivity of the nodes inside an observation region $\mathcal{R}$. In this paper, we will refer to this as the *observation model*. In the literature, other models for finite networks include: i) the *restriction model*, where the network graph is strictly limited to a finite square, with no nodes outside the square (e.g., [39]), and ii) the *toroidal model*, where the network graph is defined over a torus (e.g., [40]). The main advantage of the observation and toroidal models is their homogeneity, since they eliminate boundary effects associated with the restriction model, leading to mathematically more elegant results.

[22]We say that an event occurs "asymptotically almost surely" (a.a.s.) if its probability approaches one as $\lambda_\ell \to \infty$.

*Definition 4.4 (Asymptotic In-Connectivity):* A legitimate node $x \in \Pi_\ell \cap \mathcal{R}$ is *asymptotically in-connected* with respect to a region $\mathcal{R}$ with area $A$ if $\lim_{\lambda_\ell \to \infty} p_{\text{in-con}} = 1$, for any $\lambda_{\text{e}} > 0$ and $A > 0$.[23]

The following theorem characterizes the asymptotic out-connectivity in the $i\mathcal{S}$-graph.

*Theorem 4.1 (Asymptotic Out-Connectivity):* For the Poisson $i\mathcal{S}$-graph with $\lambda_{\text{e}} > 0$ and $A > 0$, the legitimate node at the origin is asymptotically out-connected.

*Proof:* Without loss of generality, consider that a legitimate node is placed at the origin, and let the region $\mathcal{R}$ be a square of size $\sqrt{A} \times \sqrt{A}$ containing at the origin. Let us partition $\mathcal{R}$ into equal subsquares $\mathcal{S}_i$ of size $\sqrt{\frac{\log \lambda_\ell - \epsilon(\lambda_\ell)}{\lambda_\ell}} \times \sqrt{\frac{\log \lambda_\ell - \epsilon(\lambda_\ell)}{\lambda_\ell}}$, such where $\epsilon(\lambda_\ell) > 0$ is the smallest number that the total number $\frac{A\lambda_\ell}{\log \lambda_\ell - \epsilon(\lambda_\ell)}$ of subsquares is an integer.[24] This partition is depicted in Fig. 12(a). A subsquare is said to be *full* if it contains at least one legitimate node, and *empty* otherwise. The probability that a subsquare is full is $1 - e^{-\log \lambda_\ell + \epsilon(\lambda_\ell)}$, and the probability that every subsquare of $\mathcal{R}$ is full is

$$\mathbb{P}\left\{ \bigwedge_{i=1}^{\frac{A\lambda_\ell}{\log \lambda_\ell - \epsilon(\lambda_\ell)}} \mathcal{S}_i \text{ is full} \right\} = \left(1 - e^{-\log \lambda_\ell + \epsilon(\lambda_\ell)}\right)^{\frac{A\lambda_\ell}{\log \lambda_\ell - \epsilon(\lambda_\ell)}}, \tag{32}$$

where we used the fact that $\Pi_\ell$ is a Poisson process. When we take the limit $\lambda_\ell \to \infty$, it is easy to see that $\epsilon(\lambda_\ell) \to 0$ and that (32) converges to one. In other words, the described partition of $\mathcal{R}$ ensures that every subsquare $\mathcal{S}_i$ will be full a.a.s.

Next, we analyze the secure connectivity between legitimate nodes belonging to *adjacent* subsquares of $\mathcal{R}$. Recall Fig. 6, where $\mathcal{S}_1$ and $\mathcal{S}_2$ denote two adjacent squares. Using an argument analogous to Section III-D1, we know that if the $7 \times 8$-subsquare rectangle ($\mathcal{Z}(a)$ in the figure) is free of eavesdroppers, then all legitimate nodes inside $\mathcal{S}_1 \cup \mathcal{S}_2$ form a single strong component.[25] Now consider a region $\mathcal{R}_{\text{sc}} \subseteq \mathcal{R}$ constructed in the following way. For every possible pair of 27adjacent subsquares $(\mathcal{S}_i, \mathcal{S}_j)$ in $\mathcal{R}$, determine whether the associated rectangle $\mathcal{Z}(\mathcal{S}_i, \mathcal{S}_j)$ is free of eavesdroppers. If so, update $\mathcal{R}_{\text{sc}}$ such that it now becomes $\mathcal{R}_{\text{sc}} \cup \mathcal{S}_i \cup \mathcal{S}_j$. Repeat the process until there are no more pairs of adjacent subsquares. With this definition, it is possible for large enough $\lambda_\ell$ to partition the square $\mathcal{R}$ into two regions as

$$\mathcal{R} = \mathcal{R}_{\text{sc}} \cup \mathcal{R}_{\text{e}},$$

where $\mathcal{R}_{\text{e}} = \mathcal{R} \backslash \mathcal{R}_{\text{sc}}$ is simply the remaining region of $\mathcal{R}$ after $\mathcal{R}_{\text{sc}}$ is constructed as above. This partition is shown in Fig. 12(a). By construction, it is easy to see that as $\lambda_\ell$ approaches infinity (or, equivalently, the size of the subsquares $\{\mathcal{S}_i\}$ approaches zero) the following properties hold a.s.:

1) The region $\mathcal{R}_{\text{e}}$ can be decomposed into non-overlapping regions as $\mathcal{R}_{\text{e}} = \bigcup_{n=1}^{N_{\text{e}}} \mathcal{R}_{\text{e}}^{(n)}$, where $N_{\text{e}} \triangleq \Pi_{\text{e}}\{\mathcal{R}\}$ is the number of eavesdroppers inside $\mathcal{R}$, and $\mathcal{R}_{\text{e}}^{(n)} \subset \mathcal{R}$ is a square of size $7 \times 7$ subsquares centered at the $n$-th eavesdropper of $\mathcal{R}$. If $N_{\text{e}} = 0$, then $R_{\text{e}} = \oslash$.

2) The origin belongs to $\mathcal{R}_{\text{sc}}$.

3) There exists a lattice path (i.e., a path composed only of horizontal and vertical segments inside $\mathcal{R}$) between every two subsquares of $\mathcal{R}_{\text{sc}}$, and thus all legitimate nodes inside $\mathcal{R}_{\text{sc}}$ form a single strong component.

---

[23]In our study of asymptotic connectivity, it is irrelevant whether we consider the observational, restriction, or toroidal model. The reason is that, as we shall see, full connectivity is determined by the behavior of the legitimate nodes *in the vicinity of the eavesdroppers*. Therefore, when we let $\lambda_\ell \to \infty$, there exist enough legitimate nodes between the border of the region $\mathcal{R}$ and any eavesdropper, so the border effects essentially vanish before they can affect the vicinity of the eavesdroppers (and thus, full connectivity).

[24]We have explicitly indicated the dependence of $\epsilon$ on $\lambda_\ell$, and for simplicity omitted its dependence on $A$ (which will be kept fixed).

[25]Note that here we are considering the case of $\varrho = 0$, while the discussion in Section III-D1 was valid for nonzero $\varrho$ as well.

We thus conclude that the origin is a.a.s. out-connected to all legitimate nodes inside $\mathcal{R}_{\text{sc}}$. It remains to determine whether it is also out-connected to all legitimate nodes inside $\mathcal{R}_{\text{e}}$. For that purpose, we consider the behaviour of the $i\mathcal{S}$-graph in the vicinity of the $n$-th eavesdropper of $\mathcal{R}$, which we denote by $e_n$.[26] We know that a node $x_i \in \Pi_\ell \cap \mathcal{R}_{\text{e}}^{(n)}$ will be in-connected iff the corresponding Voronoi cell induced by the process $\Pi_{\text{e}} \cup \{x_i\}$ has at least another legitimate node [31]. A little reflection shows that as $\lambda_\ell \to \infty$ this Voronoi cell approaches the half-plane

$$\mathcal{H}(x_i) \triangleq \{y \in \mathbb{R}^2 : |y - x_i| < |y - e_n|\},$$

as depicted in Fig. 12(b). Now, it is easy to see that for every $x_i \in \Pi_\ell \cap \mathcal{R}_{\text{e}}^{(n)}$, there is a.a.s. at least one legitimate node inside the region $\mathcal{H}(x_i) \cap \mathcal{R}_{\text{sc}}$, and thus every such node $x_i$ has an in-connection from the strong component in $\mathcal{R}_{\text{sc}}$. This argument holds similarly for every region $\mathcal{R}_{\text{e}}^{(n)}, n = 1, \ldots, N_{\text{e}}$, and so we conclude that the origin is a.a.s. out-connected to all legitimate nodes inside $\mathcal{R}_e$, in addition to those in $\mathcal{R}_{sc}$. This is the result of the theorem and the proof is concluded. □

---

[26]In the trivial case of zero eavesdroppers in $\mathcal{R}$, the origin is out-connected to all legitimate nodes inside $\mathcal{R}$, and the theorem follows.

The following theorem characterizes the asymptotic in-connectivity in the $i\mathcal{S}$-graph.

*Theorem 4.2 (Asymptotic In-Connectivity):* For the Poisson $i\mathcal{S}$-graph with $\lambda_e > 0$ and $A > 0$, we have that

$$\lim_{\lambda \to \infty} p_{\text{in-con}} \leq 1 - \frac{6\pi}{8\pi + 3\sqrt{3}}(1 - e^{-\lambda_e A}), \tag{33}$$

i.e., the legitimate node at the origin is *not* asymptotically in-connected.

*Proof:* Consider a region $\mathcal{R}$ with area $A$, where a probe legitimate node (node 0) is placed at the origin. Let and $\Pi_\ell\{\mathcal{R}\}$ and $\Pi_e\{\mathcal{R}\}$ denote the number of nodes in $\Pi_\ell \cap \mathcal{R}$ and $\Pi_e \cap \mathcal{R}$, respectively. Consider that the event that there is at least one eavesdropper and one legitimate node in region $\mathcal{R}$, as depicted in Fig. 13. Let $\chi_1$ denote the distance between a arbitrarily selected eavesdropper $e$ and its *closest* legitimate node $x_1 \in \mathcal{R}$, i.e., $\chi_1 \triangleq |e - x_1|$. In addition, let $\mathcal{S}$ be the set of possible locations in $\mathbb{R}^2$ where a node can connect to $x_1$, given that $x_1$ is the closest legitimate node to $e$, i.e.,

$$\begin{aligned}\mathcal{S} &\triangleq \{x \in \mathbb{R}^2 : \overrightarrow{x_1 x} \text{ is possible } \wedge |x - e| > \chi_1\} \\ &= \{x \in \mathbb{R}^2 : |x - x_1| < \chi_1 \wedge |x - e| > \chi_1\} \\ &= \mathcal{B}_{x_1}(\chi_1) \backslash \mathcal{B}_e(\chi_1),\end{aligned}$$

and is shown in Fig. 13. We now define the event $E_3 \triangleq \{\Pi_e\{\mathcal{R}\} \geq 1 \wedge \Pi_\ell\{\mathcal{R}\} \geq 1 \wedge \Pi_\ell\{\mathcal{S}\} = 0\}$. Note that if there are no legitimate nodes inside $\mathcal{S}$, then $x_1$ is *out-isolated*, and the origin is *not* fully in-connected, i.e., $E_3 \subseteq \overline{E_2}$. As a consequence, we have that

$$\mathbb{P}\{\overline{E_2}\} \geq \mathbb{P}\{E_3\}$$

or

$$1 - p_{\text{in-con}} \geq \mathbb{P}\{\Pi_e\{\mathcal{R}\} \geq 1 \wedge \Pi_\ell\{\mathcal{R}\} \geq 1 \wedge \Pi_\ell\{\mathcal{S}\} = 0\},$$

which can be manipulated as follows

$$\begin{aligned}p_{\text{in-con}} &\leq 1 - \mathbb{P}\{\Pi_e\{\mathcal{R}\} \geq 1 \wedge \Pi_\ell\{\mathcal{R}\} \geq 1 \wedge \Pi_\ell\{\mathcal{S}\} = 0\} \\ &= 1 - (1 - e^{\lambda_E A}) \cdot \mathbb{P}\{\Pi_\ell\{\mathcal{R}\} \geq 1 \wedge \Pi_\ell\{\mathcal{S}\} = 0\},\end{aligned}$$

where we used the fact that $\Pi_\ell$ and $\Pi_\mathrm{e}$ are independent processes. We now take limits as $\lambda_\ell \to \infty$ on both sides while keeping $\lambda_\mathrm{e}$ and $A$ fixed. For the purposes of determining $\mathbb{P}\{\Pi_\ell\{\mathcal{R}\} \geq 1 \wedge \Pi_\ell\{\mathcal{S}\} = 0\}$, letting $\lambda_\ell \to \infty$ with $A$ fixed is equivalent to letting $A \to \infty$ with $\lambda_\ell$ fixed. In such limiting regime of an *infinite area* Poisson process, the event $\{\Pi_\ell\{\mathcal{R}\} \geq 1\}$ occurs a.s., and $\lim_{\lambda \to \infty} \mathbb{P}\{\Pi_\ell\{\mathcal{R}\} \geq 1 \wedge \Pi_\ell\{\mathcal{S}\} = 0\} = \mathbb{P}\{\Pi_\ell\{\mathcal{S}\} = 0\}$. Then,

$$\lim_{\lambda_\ell \to \infty} p_{\mathrm{in-con}} \leq 1 - (1 - e^{\lambda_\mathrm{e} A}) \cdot \mathbb{P}\{\Pi_\ell\{\mathcal{S}\} = 0\}. \tag{34}$$

To determine $\mathbb{P}\{\Pi_\ell\{\mathcal{S}\} = 0\}$, we use two facts: 1) when conditioned on $\chi_1$, the area of $\mathcal{S}$ is equal to $\pi \chi_1^2 \left(\frac{1}{3} + \frac{\sqrt{3}}{2\pi}\right)$; and 2) when $\lambda_\ell \to \infty$, the boundary effects vanish, and the RV $\zeta \triangleq \chi_1^2$ becomes exponentially distributed with rate $\pi \lambda_\ell$. Then,

$$\begin{aligned} \mathbb{P}\{\Pi_\ell\{\mathcal{S}\} = 0\} &= \mathbb{E}_{\chi_1}\{\mathbb{P}\{\Pi_\ell\{\mathcal{S}\} = 0 | \chi_1\}\} \\ &= \mathbb{E}_{\chi_1}\left\{\exp\left(-\lambda \pi \chi_1^2 \left(\frac{1}{3} + \frac{\sqrt{3}}{2\pi}\right)\right)\right\} \\ &= \int_0^\infty \exp\left(-\lambda_\ell \pi \zeta \left(\frac{1}{3} + \frac{\sqrt{3}}{2\pi}\right)\right) \pi \lambda_\ell \exp(-\pi \lambda_\ell \zeta) d\zeta \\ &= \frac{6\pi}{8\pi + 3\sqrt{3}}. \end{aligned}$$

With this result, (34) becomes

$$\lim_{\lambda_\ell \to \infty} p_{\mathrm{in-con}} \leq 1 - \frac{6\pi}{8\pi + 3\sqrt{3}} (1 - e^{-\lambda_\mathrm{e} A}),$$

which is the bound in (33). Thus, the legitimate node at the origin is not asymptotically in-connected, and the proof is concluded. $\square$

The theorem has the following intuitive explanation. Consider $\lambda_\ell$ (or $A$) large enough that border effects can be ignored. Given that exactly one eavesdropper occurs inside region $\mathcal{R}$, there is a constant probability $\mathbb{P}\{\Pi_\ell\{\mathcal{S}\} = 0\} = \frac{6\pi}{8\pi + 3\sqrt{3}} \approx 0.62$ that the legitimate node closest to the eavesdropper is out-isolated, and this probability does not decrease with $\lambda_\ell$. In fact, when $\lambda_\ell$ increased, the area of $\mathcal{S}$ decreases in such a way that $\mathbb{P}\{\Pi_\ell\{\mathcal{S}\} = 0\}$ remains constant. As a result, regardless of how large $\lambda_\ell$ is made, there is a constant probability of $\approx 0.62$ that the nearest node is out-isolated, and therefore a positive probability that the origin is *not* in-connected.

Theorems 4.1 and 4.2 clearly show that increasing the density $\lambda_\ell$ of legitimate nodes is an effective way to improve the full out-connectivity, in the sense that the corresponding probability approaches one. However, the probability of full in-connectivity *cannot* be made arbitrarily close to one by increasing $\lambda_\ell$. In essence, full (in or out) connectivity is determined by the behavior of the legitimate nodes in the vicinity of the eavesdroppers. It is more likely that a legitimate node in such vicinity is *locally* in-connected than out-connected [31, Property 3.3], which is reflected in the fact that the origin achieves full out-connectivity a.a.s., but not full in-connectivity. Operationally, this means a node can a.a.s. *transmit* secret messages to all the nodes in a finite region $\mathcal{R}$, but cannot a.s.s. *receive* secret messages from all the nodes in $\mathcal{R}$.

Recall that for the study of full connectivity, we considered only the simplest scenario of $\varrho = 0$. Using a coupling argument similar to the proof of Lemma 3.3, it is easy to show that the probabilities $p_{\text{out-con}}(\varrho)$ and $p_{\text{in-con}}(\varrho)$ are decreasing functions of $\varrho$. In other words, the case of $\varrho = 0$ represents of the most favorable scenario in terms of full connectivity.

## C. Full Connectivity: Finite Regime

We now attempt to characterize full connectivity for a finite density of legitimate nodes. We start with the simple observation that if node $0$ is fully-out connected, then there are no in-isolated nodes in $\mathcal{R}$. Then, we immediately obtain an upper bound for $p_{\text{out-con}}$ as

$$p_{\text{out-con}} \leq \mathbb{P}\{\text{no in-isolated nodes in } \mathcal{R}\}. \tag{35}$$

We would like to express the right-hand side in terms of the individual in-isolation probability determined in [31, eq. (13)]. In general, this is non-trivial because the in-isolation events for different nodes are statistically dependent. For example, if legitimate node $x_A$ is in-isolated and node $x_B$ is close to $x_A$, then it is most likely that $x_B$ is also in-isolated. Full-connectivity has been previously studied in the case of the Poisson Boolean model for unsecured wireless networks.[27] For such scenario, it has been shown in [35], [55], [56] that as the average node degree $\pi \lambda r_{\max}^2$ becomes large, two phenomena are observed: 1) the isolation events for different nodes become almost independent; and 2) $\mathbb{P}\{\text{full connectivity}\} \approx \mathbb{P}\{\text{no isolated nodes}\}$, i.e., a bound analogous to (35) becomes tight. These two facts imply that for the Poisson Boolean model, the $\mathbb{P}\{\text{no isolated nodes}\}$ is both a simple and accurate analytical approximation for $\mathbb{P}\{\text{full connectivity}\}$, when $\pi \lambda r_{\max}^2 \to \infty$.

---

[27]The Poisson Boolean model is an undirected model where each node can establish wireless links to all nodes within a fixed connectivity range $r_{\max}$, but to no other.

We now investigate under which conditions similar phenomena occur in the $i\mathcal{S}$-graph. For that purpose, we introduce the following definition:

$$\widetilde{p}_{\text{out-con}} \triangleq \mathbb{E}_{N_\mathcal{R}}\{(1 - p_{\text{in-isol}})^{N_\mathcal{R}}\}, \tag{36}$$

where $N_\mathcal{R} = \Pi_\ell\{\mathcal{R}\}$ is the random number of legitimate nodes inside the region $\mathcal{R}$ (excluding the probe node at the origin). The quantity $\widetilde{p}_{\text{out-con}}$ represents the probability that none of the $N_\mathcal{R}$ legitimate nodes are in-isolated, under the approximation that the in-isolation events are *independent* and have the same probability $p_{\text{in-isol}}$ given in [31, eq. (13)]. As we will show later, this quantity can serve as a good approximation of $p_{\text{out-con}}$, with the advantage that it only depends on local characteristics (the isolation probabilities) of the $i\mathcal{S}$-graph and is analytically tractable. This can be shown by rewriting (36) as

$$\begin{aligned}
\widetilde{p}_{\text{out-con}} &= \sum_{n=0}^{\infty} \frac{(\lambda_\ell A)^n \exp(-\lambda_\ell A)}{n!}(1 - p_{\text{in-isol}})^n \\
&= \exp(-\lambda_\ell A\, p_{\text{in-isol}}) \underbrace{\sum_{n=0}^{\infty} \frac{[\lambda_\ell A(1 - p_{\text{in-isol}})]^n \exp(-\lambda_\ell A(1 - p_{\text{in-isol}}))}{n!}}_{=1} \\
&= \exp(-\lambda_\ell A\, p_{\text{in-isol}}) \\
&= \exp\left(-\lambda_\ell A\, \mathbb{E}\left\{e^{-\frac{\lambda_\ell}{\lambda_e}\widetilde{A}}\right\}\right),
\end{aligned} \tag{37}$$

where $\widetilde{A}$ is the (random) area of a typical Voronoi cell induced by a unit-density Poisson process. Here, we used the expression for $p_{\text{in-isol}}$ in [31, eq. (13)].

For the case of full in-connectivity, we can proceed in a completely analogous way to write

$$p_{\text{in-con}} \leq \mathbb{P}\{\text{no out-isolated nodes in } \mathcal{R}\}, \tag{38}$$

and

$$\begin{aligned}
\widetilde{p}_{\text{in-con}} &\triangleq \mathbb{E}_{N_\mathcal{R}}\{(1 - p_{\text{out-isol}})^{N_\mathcal{R}}\} \\
&= \exp(-\lambda_\ell A\, p_{\text{out-isol}}) \\
&= \exp\left(-A\frac{\lambda_\ell \lambda_e}{\lambda_\ell + \lambda_e}\right),
\end{aligned} \tag{39}$$

where we used the expression for $p_{\text{out-isol}}$ in [31, Eq. (18)].

Furthermore, according to [31, Property 3.3], we know that $p_{\text{in}-\text{isol}} < p_{\text{out}-\text{isol}}$ for $\lambda_\ell > 0$ and $\lambda_\text{e} > 0$, and therefore

$$\widetilde{p}_{\text{out}-\text{con}} > \widetilde{p}_{\text{in}-\text{con}}.$$

As as result, in the regime where $\widetilde{p}_{\text{in}-\text{con}}$ and $\widetilde{p}_{\text{out}-\text{con}}$ closely approximate $p_{\text{in}-\text{con}}$ and $p_{\text{out}-\text{con}}$, respectively, then $p_{\text{out}-\text{con}}$ will be typically larger than $p_{\text{in}-\text{con}}$. Intuitively, it is *easier* for an individual node to be *locally in-connected* than out-connected, and this fact is reflected in the global connectivity properties of the $i\mathcal{S}$-graph, in the sense that is *easier* for the origin to be *fully out-connected* (reach all nodes) than fully in-connected (be reached by all nodes).

## D. Simulation Results

We resort to Monte Carlo simulations to study full-connectivity in the $i\mathcal{S}$-graph, and in particular the accuracy of the approximations introduced in the previous section. In our environment, we define the region $\mathcal{R} = [-5, 5]\,\text{m} \times [-5, 5]\,\text{m}$ with area $A = 100\,\text{m}^2$. We place $\Pi_\ell\{\mathcal{R}\} \sim \mathcal{P}(\lambda_\ell A)$ legitimate nodes and $\Pi_\text{e}\{\mathcal{R}\} \sim \mathcal{P}(\lambda_\text{e} A)$ legitimate nodes inside $\mathcal{R}$, uniformly and independently. The $i\mathcal{S}$-graph $G = \{\Pi_\ell, \mathcal{E}\}$ is then established using as edge set

$$\mathcal{E} = \left\{ \overrightarrow{x_i x_j} : d(x_i, x_j) < d(x_i, e^*) \quad e^* = \operatorname*{argmin}_{e_k \in \Pi_\text{e}} d(x_i, e_k) \right\}, \tag{40}$$

where $d(\cdot, \cdot)$ is a toroidal distance metric, similarly to Section III-E. As discussed in Footnote 21, our definitions of full connectivity imply that legitimate nodes outside the observation region $\mathcal{R}$ can act as relays to connect other legitimate nodes inside $\mathcal{R}$. Thus, an Euclidean metric $|x_i - x_j|$ over the finite region $\mathcal{R}$ would again give rise to boundary effects, so we use a toroidal distance metric to minimize such effects in the simulations. After the $i\mathcal{S}$-graph is established, we check whether: (a) there are any (in or out) isolated nodes, and (b) the node at the origin is fully (in or out) connected. Repeating the procedure over an ensemble of $20,000$ spatial realizations of $\Pi_\ell$ and $\Pi_\text{e}$, we calculate the various probabilities of interest.

Figure 14 considers full out-connectivity, comparing three different probabilities as a function of $\lambda_\text{e}$ and $\lambda_\ell$:

- the simulated $\mathbb{P}\{\text{no in-isolated nodes in } \mathcal{R}\}$, which is an upper bound for $p_{\text{out}-\text{con}}$ as given in (35);
- the analytical $\widetilde{p}_{\text{out}-\text{con}}$, whose expression is given in (37);

- the simulated probability of full out-connectivity, $p_{\text{out-con}}$.

From the plots, we observe that the analytical curve $\widetilde{p}_{\text{out-con}}$ approximates $p_{\text{out-con}}$ surprisingly well for all $\lambda_\ell$ and $\lambda_e$, considering the strong approximations associated with $\widetilde{p}_{\text{out-con}}$. Furthermore, the approximation becomes tight in the extremes ranges where $\lambda_\ell A \to \infty$ or $\lambda_e A \to 0$ (i.e., $p_{\text{out-con}} \approx 1$). This corresponds to a regime of practical interest where is desirable to operate the network, in the sense that secure out-connectivity is achieved with probability very close to one.

Figure 15 is analogous to Fig. 14, but for the case of full in-connectivity. It compares $\mathbb{P}\{\text{no out-isolated nodes in } \mathcal{R}\}$, $p_{\text{in-con}}$, and $\widetilde{p}_{\text{in-con}}$, as a function of $\lambda_e$ and $\lambda_\ell$. We observe the approximation of $p_{\text{in-con}}$ by $\widetilde{p}_{\text{in-con}}$ becomes tight when $\lambda_e A \to 0$ (i.e., $p_{\text{in-con}} \approx 1$), but *not* when $\lambda_\ell A \to \infty$, unlike what happens for full out-connectivity. The difference in the behavior of $p_{\text{out-con}}$ and $p_{\text{in-con}}$ as $\lambda_\ell \to \infty$ was described in Section IV-B.

In general, based on the simulations we conclude that $\widetilde{p}_{\text{out-con}}$ and for $\widetilde{p}_{\text{in-con}}$ are fairly good approximations for the corresponding probabilities of full connectivity, for a wide range of parameters. The main advantage is that $\widetilde{p}_{\text{out-con}}$ and for $\widetilde{p}_{\text{in-con}}$ only depend on the *local* characterization of the network, namely on the isolation probabilities, and thus lead to simple analytical expressions which can be used to infer about the *global* behaviour of the network. In particular, they are simple enough to be used in first-order dimensioning of the system, providing the network designer with valuable insights on how $p_{\text{out-con}}$ and $p_{\text{in-con}}$ vary with the parameters $\lambda_\ell$, $\lambda_e$, and $A$.

## V. Conclusion

The $i\mathcal{S}$-graph captures the connections that can be established with MSR exceeding a threshold $\varrho$, in a large-scale networks. In [31], we characterized the *local properties* of the $i\mathcal{S}$-graph, including the degrees and MSR of a typical node with respect to its neighbours. In this paper, we build on that work and analyze the *global properties* of the $i\mathcal{S}$-graph, namely percolation on the infinite plane, and connectivity on a finite region. Interestingly, some local metrics such as the isolation probability, although quite simple to derive, are able to provide insights into the more complex phenomena such as global connectivity.

We first characterized percolation of the Poisson $i\mathcal{S}$-graph on the infinite plane. We showed that each of the four components of the $i\mathcal{S}$-graph (in, out, weak, and strong) experiences a phase transition at some nontrivial critical density $\lambda_c^\diamond$ of legitimate nodes. Operationally, this is important because it implies that long-range communication over multiple hops is still feasible when a secrecy constraint is present. We proved that percolation can occur for any prescribed infimum secrecy rate $\varrho$ satisfying $\varrho < \varrho_{\max} = \log_2\left(1 + \frac{P \cdot g(0)}{\sigma^2}\right)$, as long as the density of legitimate nodes is made large enough. This implies that for unbounded path loss models, percolation can occur for *any* arbitrarily large secrecy requirement $\varrho$, while for bounded models the desired $\varrho$ may be too high to allow percolation. Our results also show that as long as $\varrho < \varrho_{\max}$, percolation can be achieved even in cases where the eavesdroppers are arbitrarily dense, by making the density of legitimate nodes large enough.

Using Monte Carlo simulations, we obtained estimates for the critical densities $\lambda_c^\diamond$. In the case of $\varrho = 0$, for example, we estimated that if the density of eavesdroppers is larger than roughly $30\%$ that of the legitimate nodes, long-range communication in the weak $i\mathcal{S}$-graph is completely disrupted, in the sense that no infinite cluster arises. In the strong $i\mathcal{S}$-graph, we estimated this fraction to be about $16\%$. For a larger secrecy requirement $\varrho$, an even more modest fraction of attackers is enough to disrupt the network.

Besides considering the existence of an unbounded component on the infinite plane, we also analyzed the existence of a fully-connected $i\mathcal{S}$-graph on a finite region. Specifically, we characterized the asymptotic behavior of secure full connectivity for a large density $\lambda_\ell$ of legitimate nodes. In particular, we showed $p_{\text{out−con}}$ approaches one as $\lambda_\ell \to \infty$, and therefore full out-connectivity can be improved as much as desired by deploying more legitimate nodes. Full in-connectivity, however, remains bounded away from one, regardless of how large $\lambda_\ell$ is made. Operationally, this means a node can a.a.s. *transmit* secret messages to all the nodes in a finite region $\mathcal{R}$, but cannot a.s.s. *receive* secret messages from all the nodes in $\mathcal{R}$.

We derived simple expressions that closely approximate $p_{\text{out−con}}$ and $p_{\text{in−con}}$ for a finite density $\lambda_\ell$ of legitimate nodes. The advantage of these approximate expressions is that they only depend on the *local* characterization of the network, namely on the isolation probabilities, and thus lead to simple analytical expressions which can be used to infer about the *global* behaviour of the network. In particular, our expressions show that typically $p_{\text{out−con}} > p_{\text{in−con}}$, i.e., it is easier for a node to be fully out-connected (reach all nodes) than fully in-connected (be reached by all nodes). Our expressions explicitly show that this fact can be directly explained in terms of the *local connectivity*: it is easier for an individual node to be locally in-connected than out-connected, and this is reflected in the behaviour of global connectivity described above. Using Monte Carlo simulations, we showed that the approximate expressions are surprisingly accurate for a wide range of densities $\lambda_\ell$ and $\lambda_e$.

We are hopeful that further efforts in combining stochastic geometry with information-theoretic principles will lead to a more comprehensive treatment of wireless security.

## APPENDIX A
## PROOF OF LEMMA 3.3

*Proof:* In what follows, we use a coupling argument. For fixed parameters $\lambda_e$ and $\varrho$, we begin with an $i\mathcal{S}$-graph $G(\lambda_{\ell,2})$ whose underlying process $\Pi_\ell$ has density $\lambda_{\ell,2}$. We then thin this process by keeping each point of $\Pi_\ell$ with probability $\frac{\lambda_{\ell,1}}{\lambda_{\ell,2}}$ where $\lambda_{\ell,1} \leq \lambda_{\ell,2}$, such that when a point is removed, all its in- and out-connections are also removed. Because of the thinning property [46, Section 5.1], the resulting process of legitimate nodes has density $\lambda_{\ell,1}$, and we have therefore obtained a valid new $i\mathcal{S}$-graph $G(\lambda_{\ell,1})$, with the same parameters $\lambda_e$ and $\varrho$ as before. By construction, the two graphs $G(\lambda_{\ell,1})$ and $G(\lambda_{\ell,2})$ are coupled in such a way that $\mathcal{K}^\diamond_{\lambda_{\ell,1}}(0) \subseteq \mathcal{K}^\diamond_{\lambda_{\ell,2}}(0)$. As a result, the event $\{|\mathcal{K}^\diamond_{\lambda_{\ell,1}}(0)| = \infty\}$ implies that $\{|\mathcal{K}^\diamond_{\lambda_{\ell,2}}(0)| = \infty\}$, and it follows that $p^\diamond_\infty(\lambda_{\ell,1}) \leq p^\diamond_\infty(\lambda_{\ell,2})$. $\square$

## APPENDIX B
## ALTERNATIVE INTERPRETATION OF THE PERCOLATION PROBABILITY

We provide an alternative interpretation for the percolation probability $p^\diamond_\infty$, which is helpful to perform simulations of the percolation phenomenon.

*Proposition B.1:* Let $\mathcal{R}$ denote a square with dimensions $\sqrt{A} \times \sqrt{A}$, and $N^\diamond_\infty$ denote the number of legitimate nodes in $\mathcal{R}$ whose component $\mathcal{K}^\diamond(x)$ is infinite, i.e.,

$$N^\diamond_\infty \triangleq \#\{x \in \Pi_\ell \cap \mathcal{R} : |\mathcal{K}^\diamond(x)| = \infty\}, \tag{41}$$

where $\diamond \in \{\text{out}, \text{in}, \text{weak}, \text{strong}\}$. Then,

$$p^\diamond_\infty = \frac{\mathbb{E}\{N^\diamond_\infty\}}{\lambda_\ell A}. \tag{42}$$

*Proof:* Consider a partition of the square $\mathcal{R}$ into $M^2$ subsquares, $\{\mathcal{S}_i\}_{i=1}^{M^2}$. A subsquare is said to be *full* if it contains exactly one legitimate node, and *empty* otherwise. Let $I_i$ be a RV that has value 1 when $\mathcal{S}_i$ is full with some node $x$ for which $\mathcal{K}^\diamond(x)$ is infinite, and 0 otherwise. Then, we have

$$\mathbb{E}\{I_i\} = \mathbb{P}\{I_i = 1\}$$
$$= \mathbb{P}\{\mathcal{S}_i \text{ full}\} \cdot \mathbb{P}\{|\mathcal{K}^\diamond(x)| = \infty | \mathcal{S}_i \text{ full}\}$$
$$= \frac{\lambda_\ell A}{M^2} \exp\left(-\frac{\lambda_\ell A}{M^2}\right) \cdot \mathbb{P}\{|\mathcal{K}^\diamond(x)| = \infty | \mathcal{S}_i \text{ full}\}$$

Defining $I_M \triangleq \sum_{i=1}^{M^2} I_i$, we see that $I_M$ approaches $N_\infty^\diamond$ a.s. as $M \to \infty$. Thus, we can write

$$\mathbb{E}\{N_\infty^\diamond\} = \lim_{M \to \infty} \mathbb{E}\{I_M\}$$
$$= \lim_{M \to \infty} M^2 \mathbb{E}\{I_i\}$$
$$= \lambda_\ell A p_\infty^\diamond.$$

This is the result in (42), and the proof is complete. □

The proposition suggests an alternative interpretation for the percolation probability $p_\infty^\diamond$: although it was defined as the probability that a given node $x$ has an infinite component $\mathcal{K}^\diamond(x)$, it also represents the average fraction of nodes in region $\mathcal{R}$ for which the component $\mathcal{K}^\diamond(x)$ is infinite.

## ACKNOWLEDGEMENTS

The authors would like to thank J. N. Tsitsiklis, V. K. Goyal, and W. Suwansantisuk for their helpful suggestions.

| Symbol | Usage |
|---|---|
| $\mathbb{E}\{\cdot\}$ | Expectation operator |
| $\mathbb{P}\{\cdot\}$ | Probability operator |
| $H(X)$ | Entropy of $X$ |
| $\Pi_\ell = \{x_i\}, \Pi_\mathrm{e} = \{e_i\}$ | Poisson processes of legitimate nodes and eavesdroppers |
| $\lambda_\ell, \lambda_\mathrm{e}$ | Spatial densities of legitimate nodes and eavesdroppers |
| $\Pi\{\mathcal{R}\}$ | Number of nodes of process $\Pi$ in region $\mathcal{R}$ |
| $N_\mathrm{in}, N_\mathrm{out}$ | In-degree and out-degree of a node |
| $\mathcal{B}_x(\rho)$ | Ball centered at $x$ with radius $\rho$ |
| $\mathcal{D}(a,b)$ | Annular region between radiuses $a$ and $b$, centered at the origin |
| $\mathbb{A}\{\mathcal{R}\}$ | Area of region $\mathcal{R}$ |
| $\mathcal{K}^\diamond(x)$ | Out, in, weak, or strong component of node $x$ |
| $p_\infty^\diamond$ | Percolation probability associated with component $\mathcal{K}^\diamond(0)$ |
| $\lambda_\mathrm{c}^\diamond$ | Critical density associated with component $\mathcal{K}^\diamond(0)$ |
| $\#S$ | Number of elements in the set $S$ |
| $\mathcal{N}(\mu, \sigma^2)$ | Gaussian distribution with mean $\mu$ and variance $\sigma^2$ |

Table I

NOTATION AND SYMBOLS.

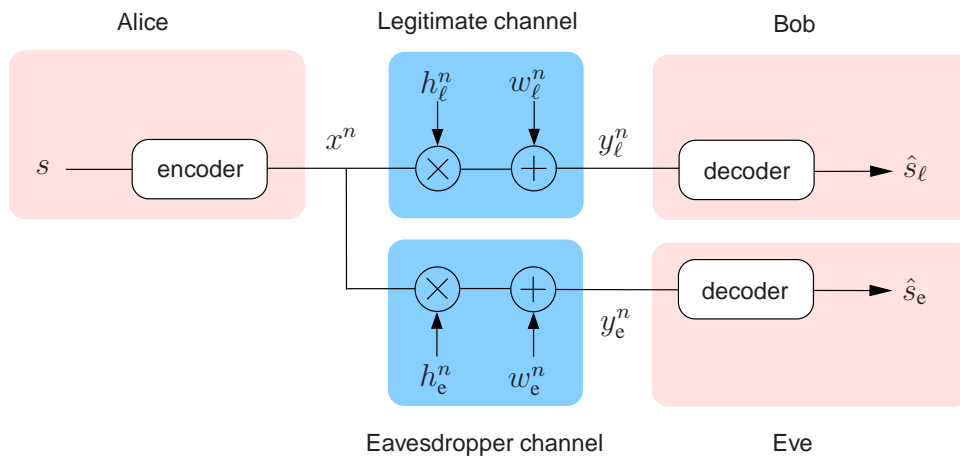

Figure 1. Wireless wiretap channel.

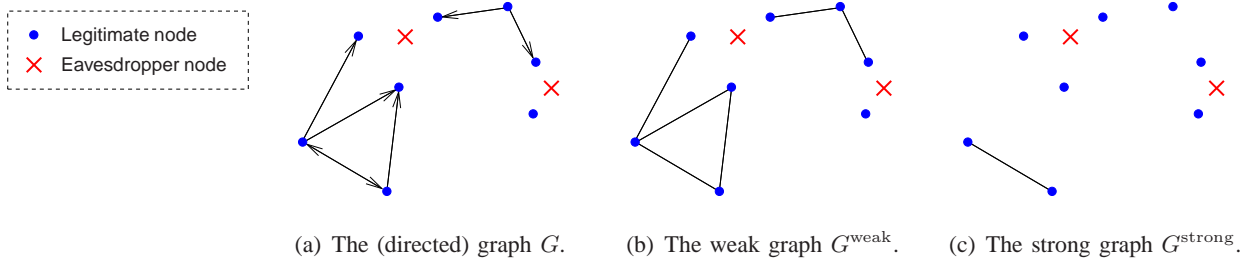

(a) The (directed) graph $G$.    (b) The weak graph $G^{\text{weak}}$.    (c) The strong graph $G^{\text{strong}}$.

Figure 2. Three different types of $i\mathcal{S}$-graphs on $\mathbb{R}^2$, considering that $\varrho = 0$ and $\sigma_\ell^2 = \sigma_e^2$.

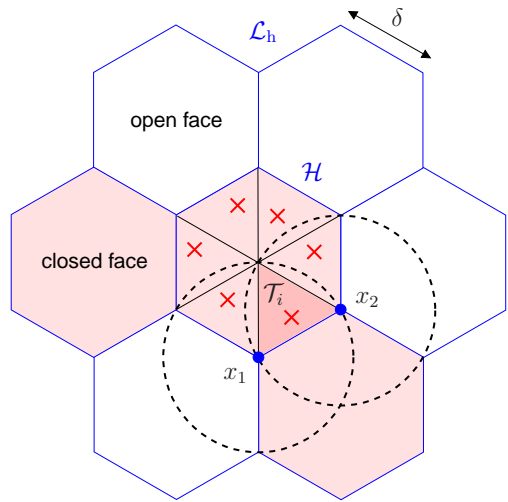

Figure 3. Conditions for a face $\mathcal{H}$ in $\mathcal{L}_\text{h}$ to be closed, according to Definition 3.1: each of the 6 triangles in $\mathcal{H}$ must have at least one eavesdropper node each, and $\mathcal{H}$ must be free of eavesdroppers.

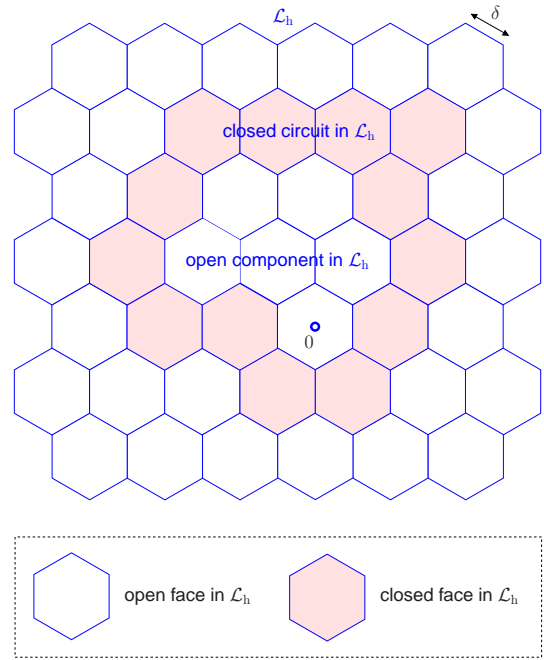

Figure 4. A finite open component at the origin, surrounded by a closed circuit.

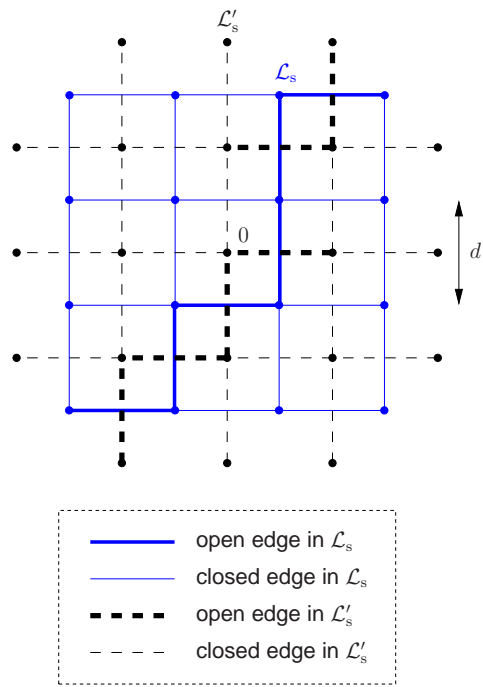

Figure 5. The lattice $\mathcal{L}_s = d \cdot \mathbb{Z}^2$ and its dual $\mathcal{L}'_s = \mathcal{L}_s + \left(\frac{d}{2}, \frac{d}{2}\right)$. We declare an edge of $\mathcal{L}'_s$ to be open iff its dual edge in $\mathcal{L}_s$ is open.

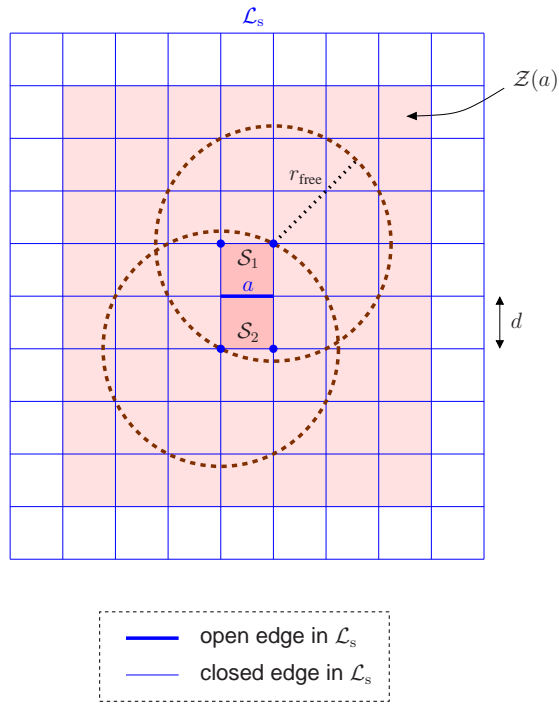

Figure 6. Conditions for an edge $a$ in $\mathcal{L}_s$ to be open, according to Definition 3.2: the squares $\mathcal{S}_1$ and $\mathcal{S}_2$ must have at least one legitimate node each, and the rectangle $\mathcal{Z}$ must be free of eavesdroppers. In general, the radius $r_{\text{free}}$—and therefore the region $\mathcal{Z}$—increase with the secrecy rate threshold $\varrho$. The figure plots the case of $\varrho = 0$.

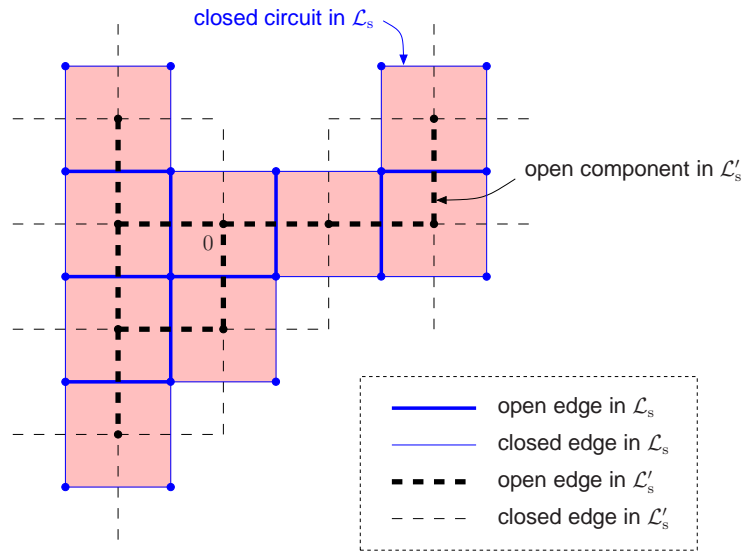

Figure 7. A finite open component at the origin, surrounded by a closed circuit in the dual lattice.

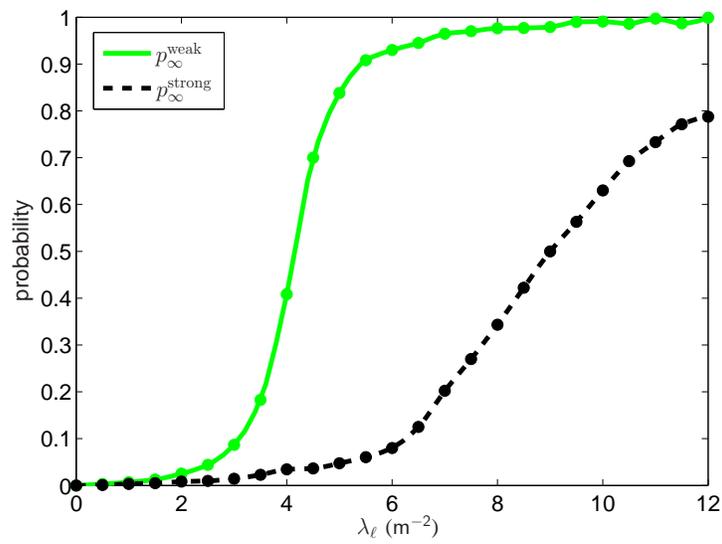

Figure 8. Simulated percolation probabilities for the weak and strong components of the $i\mathcal{S}$-graph, versus the density $\lambda_\ell$ of legitimate nodes ($\lambda_\mathrm{e} = 1\,\mathrm{m}^{-2}$, $\varrho = 0$).

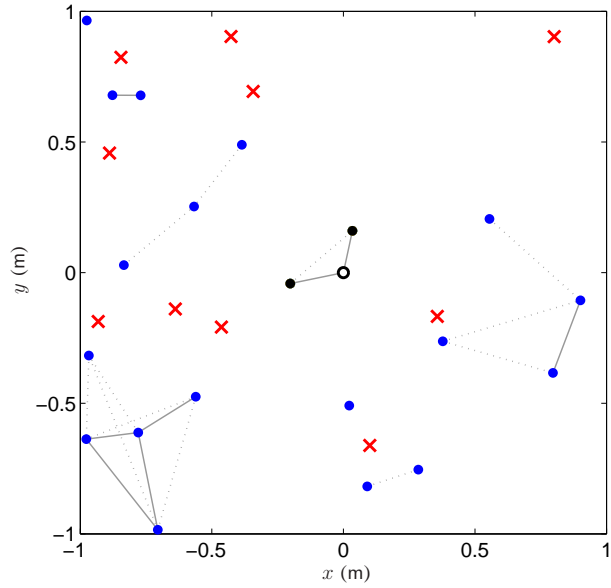

(a) Subcritical graph ($\lambda_\ell/\lambda_e = 2$).

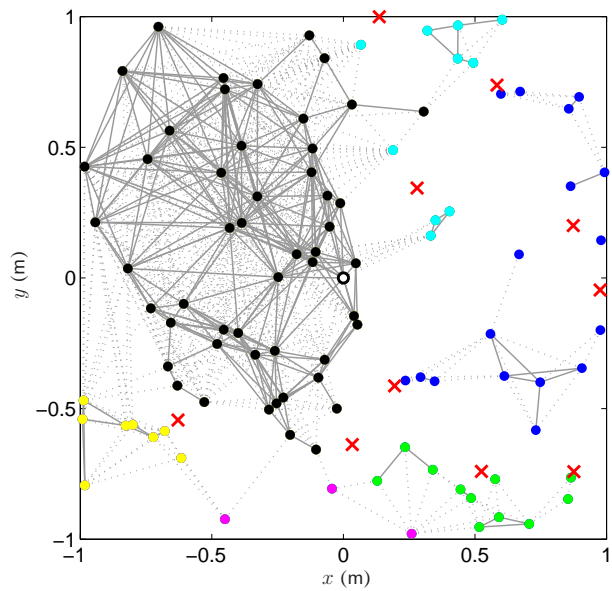

(b) Supercritical graph ($\lambda_\ell/\lambda_e = 10$).

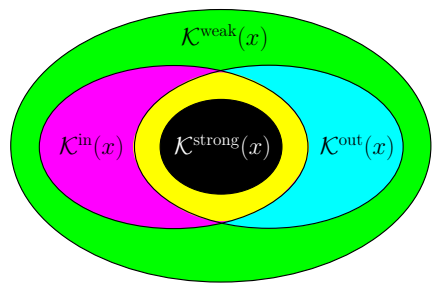

(c) Structure and color legend of the various graph components of node $x = 0$.

Figure 9. Percolation in the $i\mathcal{S}$-graph for $\varrho = 0$. The solid lines represent the edges in $G^{\text{strong}}$, while the dotted lines represent the edges in $G^{\text{weak}}$.

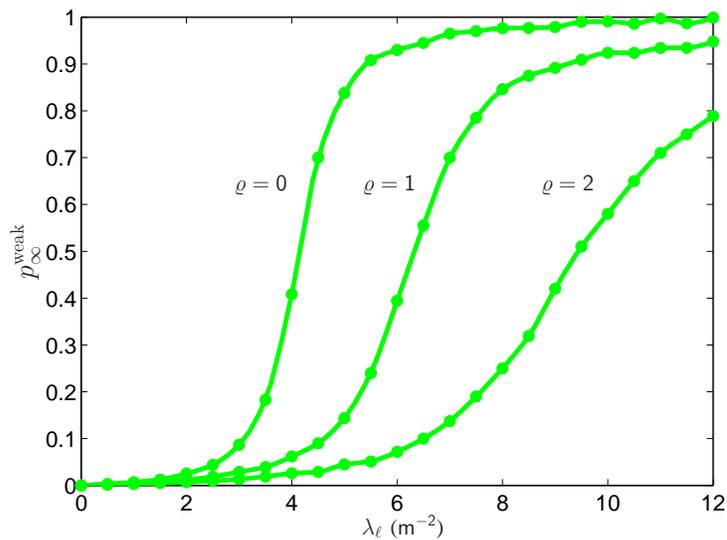

Figure 10. Effect of the secrecy rate threshold $\varrho$ on the percolation probability $p_\infty^{\text{weak}}$ ($\lambda_\text{e} = 1\,\text{m}^{-2}$, $g(r) = \frac{1}{r^{2b}}$, $b = 2$, $P_\ell/\sigma^2 = 10$).

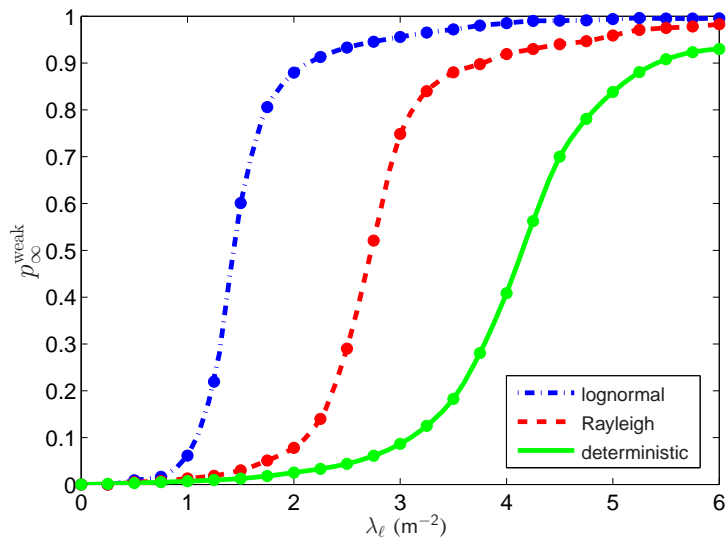

Figure 11. Effect of the wireless propagation characteristics on the percolation probability $p_\infty^{\text{weak}}$ ($\lambda_\text{e} = 1\,\text{m}^{-2}$, $\varrho = 0$, $\sigma_{\text{dB}} = 10$).

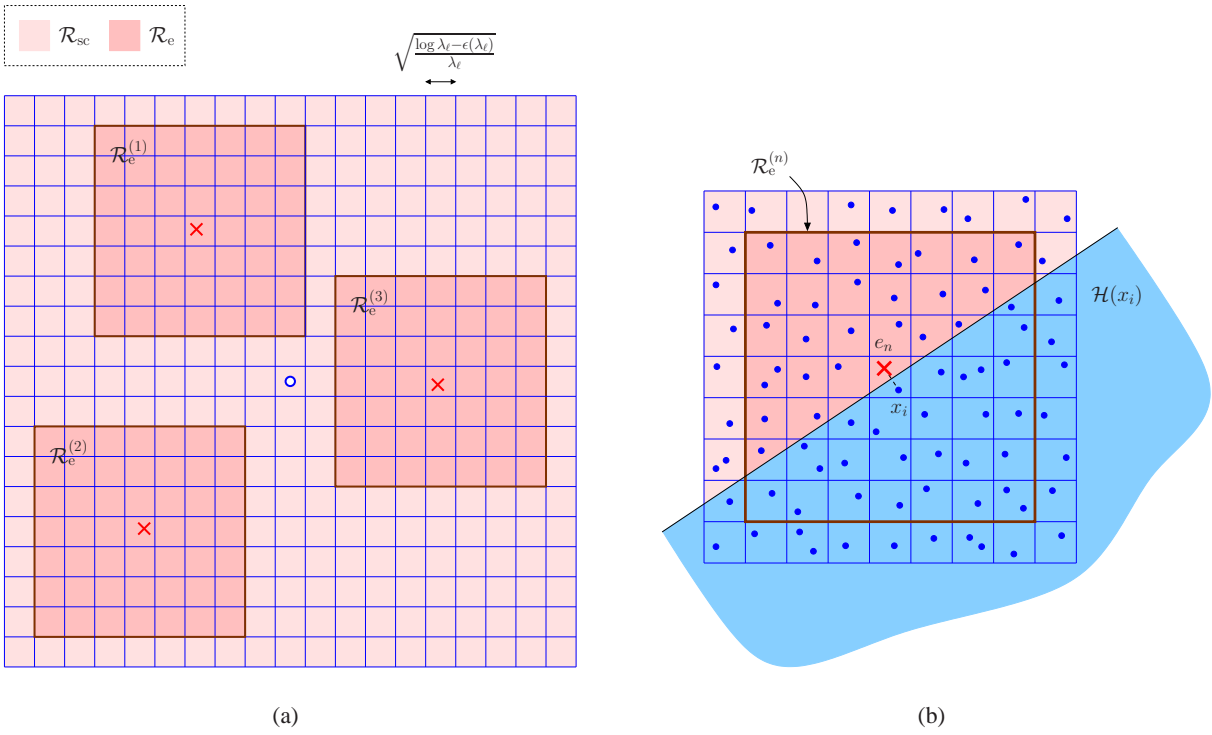

Figure 12. Auxiliary diagrams for proving that $\lim_{\lambda_\ell \to \infty} p_{\text{out-con}} = 1$.

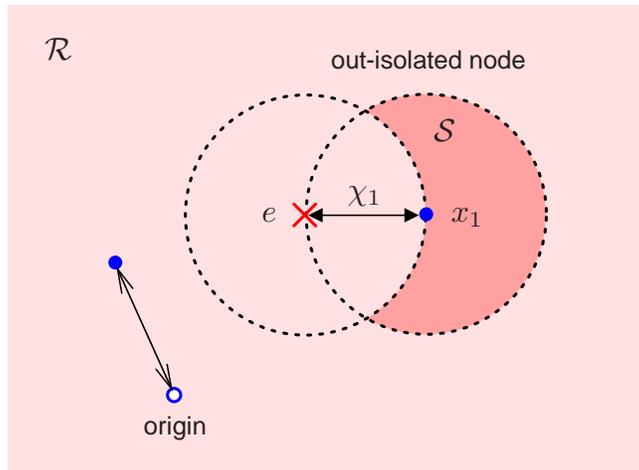

Figure 13. Auxiliary diagram for proving that $\lim_{\lambda_\ell \to \infty} p_{\text{in-con}} < 1$.

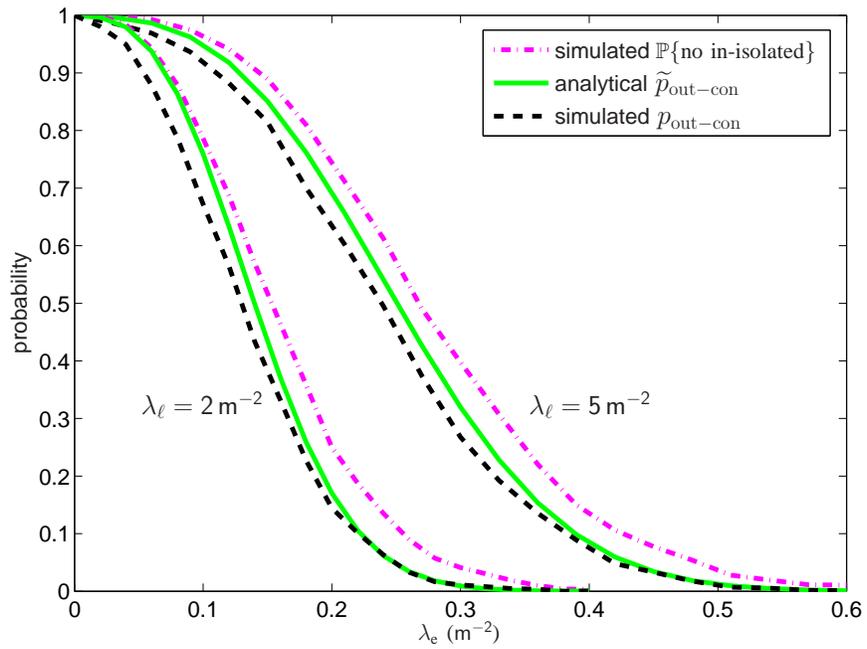

(a) Connection probabilities versus the eavesdropper density $\lambda_e$, for various values of $\lambda_\ell$.

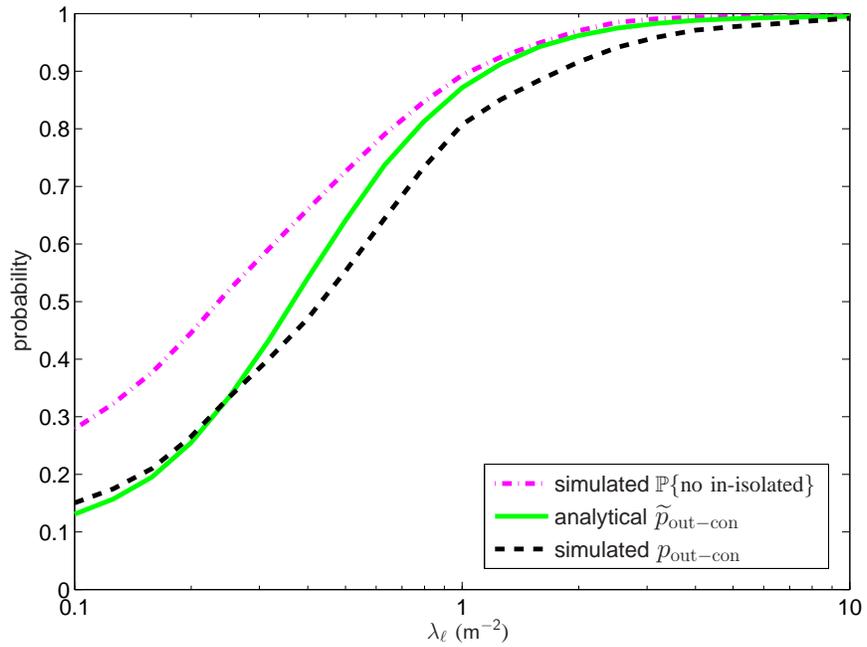

(b) Connection probabilities versus the spatial density $\lambda_\ell$ of legitimate nodes ($\lambda_e = 0.05 \, \text{m}^{-2}$).

Figure 14. Full out-connectivity in the Poisson $i\mathcal{S}$-graph ($A = 100 \, \text{m}^2$, $\varrho = 0$).

49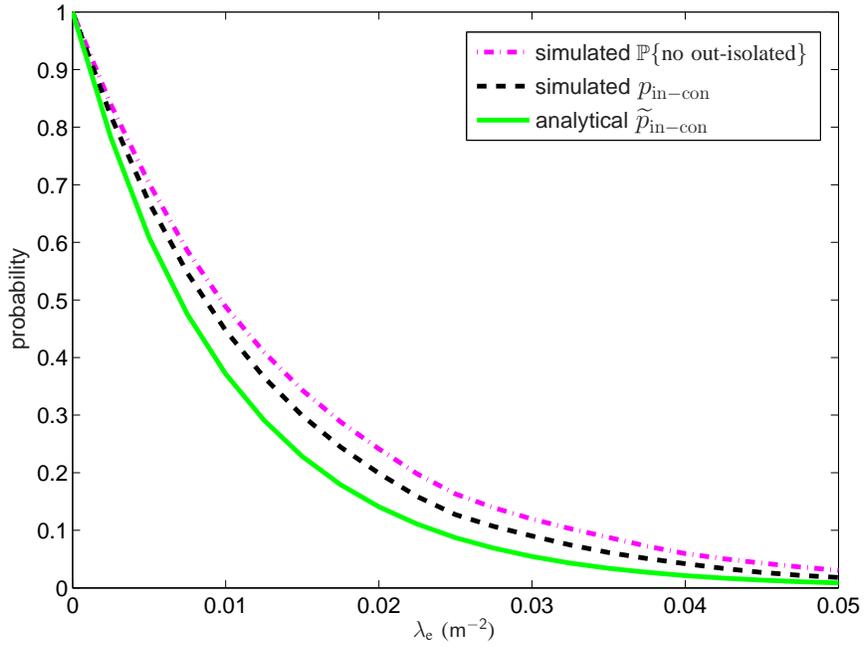

(a) Connection probabilities versus the eavesdropper density $\lambda_e$ ($\lambda_\ell = 1\,\text{m}^{-2}$).

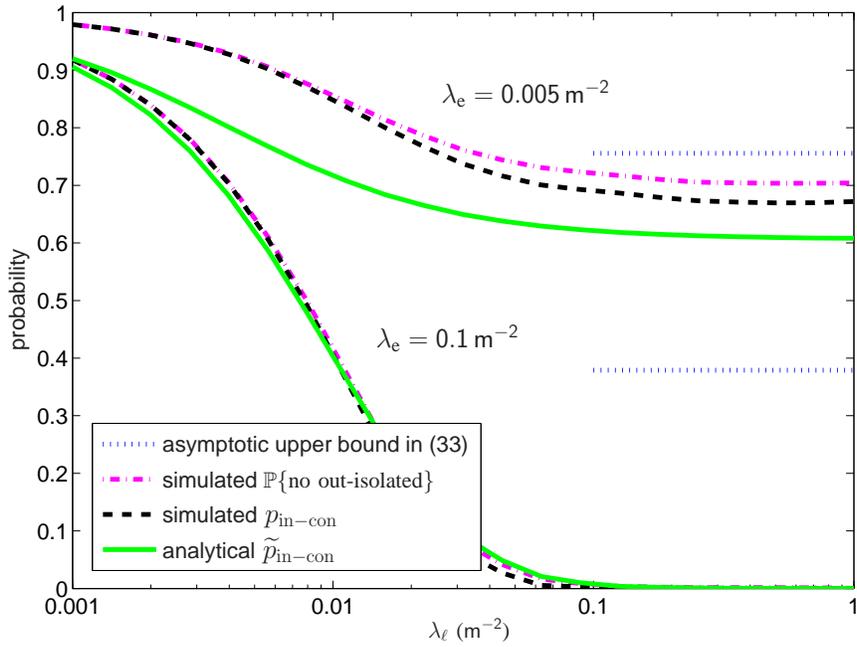

(b) Connection probabilities versus the spatial density $\lambda_\ell$ of legitimate nodes, for various values of $\lambda_e$.

Figure 15. Full in-connectivity in the Poisson $i\mathcal{S}$-graph ($A = 100\,\text{m}^2$, $\varrho = 0$).